\documentclass[prx, reprint, 9pt, %superscriptaddress,
notitlepage, nofootinbib, 
longbibliography,
floatfix]{revtex4-1}
\usepackage{gpkmac}
\usepackage{xcolor}

\newcommand{\YL}[1]{{\black #1}}

\begin{document}

%------------------------------------------------------------
\title{Statistical mechanics of %dynamically-generated
quantum error correcting codes}
\author{Yaodong Li}
\email{lyd@physics.ucsb.edu}
\affiliation{Department of Physics, University of California, Santa Barbara, CA 93106}
\author{Matthew P. A. Fisher}
\email{mpaf@kitp.ucsb.edu}
\affiliation{Department of Physics, University of California, Santa Barbara, CA 93106}
%\date{August 19, 2020}
\date{March 19, 2021}

\begin{abstract}
We study stabilizer quantum error correcting codes (QECC) generated under hybrid dynamics of local Clifford unitaries and local Pauli measurements in one dimension.
Building upon 1) a general formula relating the error-susceptibility of a subregion to its entanglement properties, and 2) a previously established mapping between entanglement entropies and domain wall free energies of %\st{some}
\YL{an} underlying spin model, we propose a statistical mechanical description of the QECC in terms of ``entanglement domain walls''.
Free energies of such domain walls generically feature a leading volume law term coming from its ``surface energy'', and a sub-volume law correction coming from thermodynamic entropies of its transverse fluctuations.
These are most easily accounted for by capillary-wave theory of liquid-gas interfaces, which we use as an illustrative tool. %throughout the paper.
We show that the information-theoretic decoupling criterion corresponds to a  geometric decoupling of domain walls, which further leads to the identification of the ``contiguous code distance'' of the QECC as the crossover length scale at which the energy and entropy of the domain wall are comparable.
The contiguous code distance thus diverges with the system size as the subleading entropic term of the free energy, protecting a finite code rate against local {undetectable} errors. 
We support these correspondences with numerical evidence, where we find capillary-wave theory describes many qualitative features of the QECC; we also discuss when and why it fails to do so.
\end{abstract}

\maketitle

%------------------------------------------------------------

\tableofcontents

\section{Introduction}

Quantum error correcting codes (QECC)~\cite{shor1995scheme, steane1996qec} are important constructions of quantum states that can be used for protecting information from decoherence and other types of errors. % commonly in the context of quantum computing.
%;
%it is very likely that QECCs will be building blocks of a scalable, fault-tolerant quantum computer.
A QECC encodes quantum information nonlocally, so that sufficiently local errors are detectable and reversible, % can only act on that encoded information in a detectable and reversible way,
allowing for explicit protocols to counter the errors~\cite{BDSW9604mixedstate, knill_laflamme_1997}. %, and their effects can be reversed by local operations.
Besides concrete constructions of QECCs with an intended use for quantum computation~\cite{calderbank9512good, steane9601multiple, BDSW9604mixedstate, gottesman9604hamming, calderbank1997quantum, gottesman1997thesis}, they can also occur naturally in physical contexts, e.g. in many-body quantum systems as a consequence of topological orders~\cite{kitaev1997, Levin2006}, or in quantum gravity as a consequence of the holographic principle~\cite{harlow1411bulklocality}.

Recently, in (1+1)-dimensional {``hybrid''} quantum circuits~\cite{nahum2018hybrid, nandkishore2018hybrid, li1808hybrid} that exhibit a ``measurement-driven transition''~\cite{li1901hybrid, choi2019qec, gullans1905purification,choi2019spin, andreas2019hybrid, gullans1910scalable, huse1911tripartite, fan2020selforganized, li2003cft}
between a highly-entangled phase and a disentangled one \YL{(see Sec.~\ref{sec:model})}, the notion of QECC also appears, and provides an interesting perspective~\cite{choi2019qec, gullans1905purification, choi2019spin, fan2020selforganized}.
The idea is to view the quantum states generated by the circuit dynamics as QECCs.
Indeed, in Clifford hybrid circuits, where numerical characterizations are most accessible, the states are ``stabilizer quantum error correcting codes'' in a strict sense, for which the ``code space'' changes at each time step of the circuit evolution.
Local measurements in the circuit can be correspondingly interpreted as ``local errors'', which tend to decrease the code rate, and when frequent enough, can drive the QECC through a transition from a phase where the QECC is resilient to local errors and thus retains a finite code rate,
to a phase where the ``error rate'' is so high that a finite code rate cannot be sustained.

A complementary approach, as firmly 
 established in hybrid random Haar circuits~\cite{andreas2019hybrid, choi2019spin},\footnote{We note that Refs.~\cite{andreas2019hybrid, choi2019spin} extended a mapping first obtained in Refs.~\cite{Hayden2016, nahum2017KPZ, nahum2018operator, zhou1804emergent} for random Haar unitary circuits \emph{without} measurements, where it was first pointed out that the entanglement entropy can be viewed as free energies of ``entanglement domain walls''.
This mapping has been extended in various contexts of unitary quantum dynamics~\cite{nahum2017quenched, nahum1803coarsegrained, zhou1912membrane}, and this development is independent of hybrid circuits.}
translates the measurement-driven transition into a ``conventional'' finite-temperature ordering transition by mapping to an underlying statistical mechanical (stat.~mech.) model of %\st{locally-interacting}
spins in {$(2+0)$-}dimensions \YL{with short-range interactions},
{with the temporal dimension of the circuit viewed as the second spatial dimension.\footnote{See also
Refs.~\cite{Hayden2016,vasseur2018rtn}, where a similar mapping was derived for random tensor networks.}
%\st{Within this mapping, the disorder-averaged entanglement entropy of a subregion corresponds to the free energy cost upon insertion of domain walls in the spin model, which accounts for the change of boundary conditions in that subregion.}
{Within this mapping, the disorder-averaged entanglement entropy of a subregion corresponds to the free energy cost upon a change of boundary condition in that subregion.
In the low-temperature ordered phase, this change of boundary condition requires the presence of sharp domain walls.}
This geometrical picture raises the possibility of a stat.~mech. description of {the entanglement structure, and, in turn, of} QECCs in hybrid circuits {in terms of these ``entanglement domain walls''}.
The aim of this work is to demonstrate such a description.

We focus on error correcting properties of stabilizer codes, as generated dynamically after running a random Clifford circuit into the steady state, where the circuit depth scales at most polynomially in the system size.
We will mostly focus on the case with a maximally-mixed initial state, and with the measurement rate below the transition threshold, $p < p_c$ (i.e. the mixed phase~\cite{gullans1905purification}; see Sec.~\ref{sec:model}), which was shown to have a finite code rate on relevant time scales.

We start in Sec.~\ref{sec:setting} by introducing the model using the stabilizer code formalism, and translate the circuit dynamics into their actions on the code space.
We then %\st{prove}
\YL{state} a theorem in Sec.~\ref{sec:theorem} that applies to all stabilizer codes, which equates the number of independent, undetectable (hence uncorrectable) errors supported on a subregion, with the mutual information between the subregion and the environment.

In Sec.~\ref{sec:dw_picture}, we review the domain wall picture of free energies as established analytically in Refs.~\cite{andreas2019hybrid, choi2019spin}, and numerically for Clifford circuits in Refs.~\cite{gullans1910scalable, li2003cft}.
Since the Clifford stat.~mech. model is not known at this stage, we choose to model the ``entanglement domain walls'' as the simplest type, i.e. that of the liquid-gas interface (or Ising domain walls) in the low-temperature phase, as described by what is called ``capillary-wave theory''~\cite{buff1965capillary, weeks1977capillary, mpaf1982capillary}. This simplification allows analytic calculations, and, as we shall see, quite generally captures qualitative features of entanglement domain walls.
Using the theorem, we translate certain {\it algebraic} properties of the QECC to %\st{geometrical}
{\emph{geometric}} properties of the domain walls.
In particular, the code rate is interpreted as the surface tension, and the code distance as the length scale below which the transverse, entropic fluctuations of the domain wall dominates over the surface energy.
The ``correctability'' of a subregion, as quantified by the ``decoupling principle'', translates into a geometric decoupling condition of domain walls.

In Sec.~\ref{sec:numerics}, we perform entanglement entropy calculations for a random Clifford circuit model, and demonstrate that capillary-wave theory gives a qualitatively accurate description of the results.
However, quantitative deviations from capillary-wave theory are present
in our numerics, which presumably reflects the specific nature of the entanglement domain walls within a stat.~mech. description for such Clifford circuit dynamics.

In Sec.~\ref{sec:discussion}, we discuss implications of our result, and mention several possible future directions.

\section{Model and setting \label{sec:setting}}

\subsection{The random Clifford circuit \label{sec:model}}

%------------------------------------------
\begin{figure}[b]
    \centering
    \includegraphics[width=.48\textwidth]{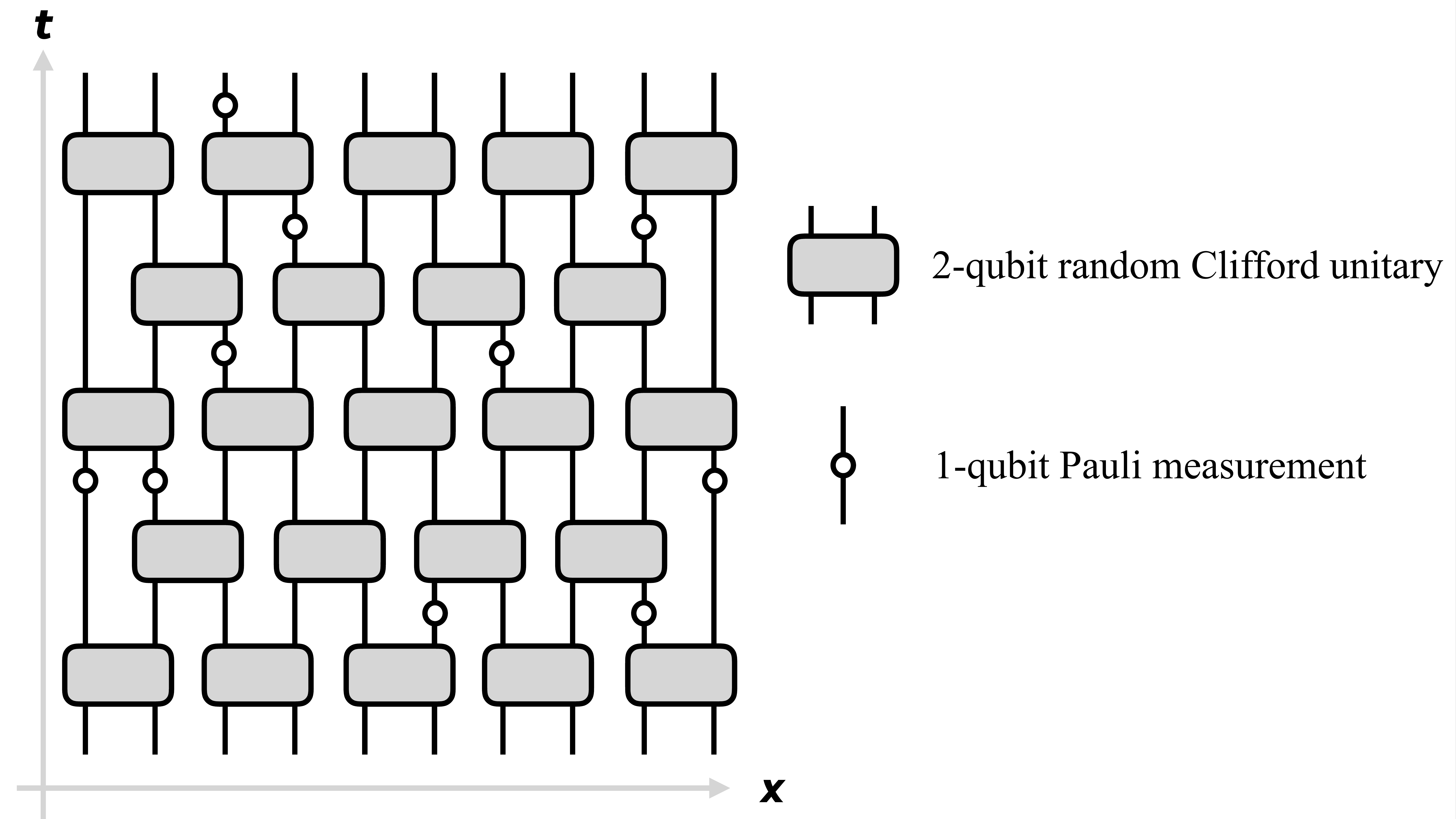}
\caption{The hybrid circuit composed of local unitaries and local measurements.
The rectangles represent random Clifford unitary gates, arranged alternatively in a ``brickwork'' fashion.
Projective measurements of single-site
Pauli operators are made between unitary layers, and at each site indepedently with probability $p < p_c$, represented by hollowed dots.
}
    \label{fig:upcircuit}
\end{figure}
%------------------------------------------

We consider ``hybrid'' circuit models~\cite{nahum2018hybrid, nandkishore2018hybrid, li1808hybrid} as shown in Fig.~\ref{fig:upcircuit}, acting on a set of  qubits $Q = \{1, \ldots, L\}$, arranged in a one-dimensional array.
The circuit is composed of nearest-neighbor unitary gates, which we restrict to the two-qubit Clifford group;\footnote{%\st{Recall that the Clifford group contains all unitaries that maps all Pauli string operators to other Pauli string operators under conjugation.}
Recall that the Clifford group contains all unitaries that maps every Pauli string operator to another %Pauli string operator
under conjugation.}
and sporadic single-qubit \YL{projective} measurements, which we restrict to be of the Pauli operators.
We focus on the quantum trajectories of the state density matrix under circuit evolution, namely,
\env{align}{
\label{eq:gate_U}
    \rho_Q &\to U \rho_Q U^\dg, &\text{under a Clifford unitary gate}, \\
\label{eq:gate_P}
    \rho_Q &\to \frac{P \rho_Q P}{\textrm{Tr}[P \rho_Q P]}, &\text{under a Pauli measurement}.
}
Here, the projection operator $P$ is given by $P = \frac{1 \pm g}{2}$, where $g$ is the Pauli operator being measured, and the plus-minus signs are the (possibly random) outcomes of the measurement.
In the case when this outcome is indeed random, we choose either outcome randomly with the corresponding probability given by Born's rule.

%By construction, the state of the circuit at any point of the time evolution is a ``stabilizer quantum error correcting code state'', or simply a ``code state''~\cite{gottesman9604hamming, gottesman9807heisenberg}, which we briefly review below.

For concreteness, we choose to sample the unitaries uniformly from the two-qubit Clifford group, and perform single-qubit measurements of probability $p$ at each time step, independently on each qubit (the ``random Clifford circuit''~\cite{li1901hybrid}).

We will focus on the maximally-mixed initial state with maximal entropy, $S(\rho_Q) = |Q| \ln 2= L \ln 2$.\footnote{
\YL{In this paper we compute the (von Neumann) entropy by taking the natural logarithm,
\env{align}{
    S(\rho) \coloneqq - \mathrm{Tr}\rho \ln \rho. \nonumber
}
This is the convention adopted in Refs.~\cite{choi2019spin, andreas2019hybrid, Hayden2016, vasseur2018rtn}, for which the equality between (average) entanglement entropies and free energies can be made (see Eq.~\eqref{eq:SA_log_Z}).
This choice of convention accounts for the extra factor of $\ln 2$ here, as well as those appearing in Eqs.~(\ref{eq:s_rho_Q_G}, \ref{eq:ell_A_I_AR}, \ref{eq:code_rate_surface_tension}) and Appendix~\ref{app:proof}.
}
}
This entropy can be equivalently thought of as the entanglement entropy between $Q$ and a ``reference system'', $R$, where $Q$ and $R$ together holds a pure state $\ket{\Psi_{QR}}$, and $\rho_Q$ is the reduced density matrix after tracing out $R$,
\env{align}{
    \rho_Q = {\rm Tr}_R \ket{\Psi_{QR}} \bra{\Psi_{QR}} = \frac{1}{2^{|Q|}}    \mathbbm{1}_Q.
}

With these specifications, the circuit model is unambiguously defined.
Within this model, the entropy of $\rho_Q$ is a monotonically decreasing function of time, $T$, the circuit depth.
The decrease of entropy is due to measurements (Eq.~\eqref{eq:gate_P}) that try to read out some information about the state, while the unitaries (Eq.~\eqref{eq:gate_U}) ``scramble/delocalize'' the information, protecting it from being read-out by local measurements.
This competition leads to a ``purification transition'' at $p = p_c \approx 0.16$~\cite{gullans1905purification, li2003cft}, where
\env{itemize}{
\item
When $p < p_c$, the state $\rho_Q$ retains a finite density of entropy at times $T$ at most polynomial in $|Q|$ (or formally $T = O(\mathrm{poly}(|Q|))$), therefore in the ``mixed phase''.\footnote{In the case of a pure initial state~\cite{nahum2018hybrid, li1808hybrid} where this transition was first found, $p < p_c$ corresponds to a ``volume law entangled phase'', where one should be thinking in terms of the reduced (mixed) density matrix of a subsystem which holds a finite entropy density as $T \to \infty$.}
\YL{An arbitrary subset of $Q$ also has a finite density of entanglement entropy, or equivalently, the entanglement entropy has a volume law scaling.}
\item
When $p > p_c$, the entropy density drops to zero on those time scales, therefore in the ``pure phase''.
\YL{A subset of $Q$ thus has zero density of entanglement entropy, or equivalently, the entanglement entropy has an area-law scaling.}
}
We will primarily restrict our attention to ``intermediate'' time scales, with  $T = O(\mathrm{poly}(|Q|))$, since for
$T$ exponentially large in $|Q|$, the circuit dynamics should fully purify the state, even when $p < p_c$.

As discussed below, the state $\rho_Q$ at any point of the circuit evolution can be thought of as a ``stabilizer code'', that can in principle be used for quantum error correcting purposes.
When {the state is} viewed as a QECC, the purification transition aquires a new interpretation:
it is a transition from ``good'' to ``bad'' QECCs, where the QECC has a nonzero/zero code rate (which is equal to the entropy density, see below) at relevant time scales, respectively, in the two phases.

%\YL{More to be said here? Maybe draw the phase diagram.}

%For this reason, the circuit dynamics can efficiently simulated in numerics~\cite{aaronson0406chp}.

%In the rest of this section, we briefly introduce stabilizer codes in Sec.~\ref{sec:stabilizer}, and recall several basic notions pertaining to its intended use in an error correcting context. In Sec.~\ref{sec:qecc_dynamics}, we translate the ``purification dynamics'' into the QECC language, with a focus on the phase diagram and a self-consistent condition on the code distance. In Sec.~\ref{sec:theorem}, we prove a general result for stabilizer codes that relates its error correcting capabilities to its entanglement structure.

\subsection{The stabilizer formalism \label{sec:stabilizer}}

Here we summarize several basic notions of stabilizer QECC %~\cite{gottesman1997thesis},
that are necessary for stating and using the theorem appearing towards the end of this section.
\YL{We refer the reader to Refs.~\cite{gottesman1997thesis, nielsen2010qiqc} and Appendix~\ref{app:proof} for details.}
%Most of these can be found in standard references, e.g. Ref.~\cite{gottesman1997thesis, nielsen2010qiqc}.
%Let $Q = \{1, \ldots, L\}$ be coordinates of the qubits.

A ``stabilizer group'' $\mc{S}$ is an abelian subgroup of the Pauli group on $Q$ ({denoted} $\mc{P}(Q)$) generated by $m \le |Q|$ independent and mutually commuting Pauli string operators,
\env{align}{
    \label{eq:s_def}
    & \mc{S} \nn
    =& \ld \prod_{j=1}^m \(g_j\)^{b_j} \bigg| b_j \in \{0,1\}, g_j \in \mc{P}(Q), [g_j, g_{j^\p}] = 0 \rd \nn %\  \forall 1 \le i, j \le m \rd \nn
    =& \avg{g_1, \ldots, g_m} \nn
    \equiv& \avg{\mc{G}} ,
}
where $\mc{G} = \{g_1, \ldots, g_m\}$ is called a ``generating set'' of $\mc{S}$.
The group $\mc{S}$ is abelian, where each element has order $2$, and can therefore be viewed as an $m$-dimensional vector space on $\mb{F}_2$, the isomorphism being given explicitly above in terms of the $b$-vector.

We list a few more properties that follow from the notion of a stabilizer group~\cite{gottesman1997thesis}:
%The organization here is not 
\env{enumerate}{
\item
A stabilizer group defines a ``code space'', that is, the subspace $\mc{H}_Q(\mc{S})$ of the Hilbert space $\mc{H}_Q$ on which all elements of $\mc{S}$ acts trivially.
We have
\env{align}{
    \dim \mc{H}_Q(\mc{S}) = 2^{k} \equiv 2^{|Q|-|\mc{G}|},
}
where $k \coloneqq |Q|-|\mc{G}| = L-m$ is known as the number of ``logical qubits'' encoded.

\item
The stabilizer group also defines its ``code state'', namely the maximally-mixed state on the code space.
Its density matrix is proportional to the projection operator onto the code space, and is explictly given by~\cite{Fattal2004stabilizer},
\env{align}{
    \label{eq:rho_S}
    \rho_Q(\mc{S}) = \frac{1}{2^{|Q|}} \sum_{g \in \mc{S}} g.
}
%, bearing in mind that the primary object is $\mc{S}$, which defines $\mc{H}_\mc{S}(Q)$, and $\rho_{\mc{S}}$.
As an example, the maximally-mixed state (i.e. the initial state of the circuit model in Fig.~\ref{fig:upcircuit}) is such a code state, for which the stabilizer group is empty.
Consequently, as we will show below, the state at any point of the random Clifford circuit evolution remains a code state, and therefore is a ``stabilizer QECC'' in a strict sense.
%which admits an efficient representation as well as an efficient simulation of the dynamics.

Since we will mostly be concerned with codes states as in Eq.~\eqref{eq:rho_S}, we will usually write $\rho_Q$ as a shorthand notation for $\rho_Q(\mc{S})$, where its dependence on $\mc{S}$ is implicit.

\item

A code state $\rho_Q$ as in Eq.~\eqref{eq:rho_S} has a flat spectrum, and all its R\'{e}nyi entropies are equal to~\cite{Fattal2004stabilizer}
\env{align}{
    \label{eq:s_rho_Q_G}
    (\ln 2)^{-1} S(\rho_Q) = |Q|-|\mc{G}| = k = \log_2 \dim \mc{H}_Q(\mc{S}).
}
That is, the entropy of the code state is equal to ($\ln 2$ times) the number of logical qubits.
%The extra factor of $\ln 2$ comes from the convention adopted in Eq.~\eqref{eq:SA_log_Z}.
It follows that the ``code rate'', defined to be the ratio between the number of logical and physical qubits, $\frac{k}{|Q|}$, is equal to the entropy density of $\rho_Q$ up to a factor of $\ln 2$.

\item
We recall that a ``logical operator'' is an element in $\mc{P}(Q)$ that commutes with all elements in $\mc{S}$, that is, an element of the centralizer $\mc{C}(\mc{S})$.\footnote{Throughout the paper, by $\mc{C}(\mc{S})$ we really mean the abelianized centralizer,
\env{align}{
    \mc{C}(\mc{S}) = \frac{\{g \in \mc{P}(Q) \ \big| \ [g, \mc{S}] = 0\}}{\{\pm 1, \pm i\}}. \nonumber
}
That is, we ``forget about'' the (uninteresting) coefficients / commutation relations of the logical operators, and focus on their operator contents.
This way, $\mc{C}(\mc{S})$ can be viewed as vector spaces on $\mb{F}_2$, and group homomorphisms (e.g. those in Appendix~\ref{app:proof}) can be viewed as linear maps between vector spaces.
On the other hand, we \emph{do} care about commutation relations of stabilizers (elements of $\mc{S}$).
We always require $\mc{S}$ to be abelian and hence identical to its abelianization.
Thus, $\mc{S}$ is a subgroup of the abelianized centralizer $\mc{C}(\mc{S})$, and the logical group $\mc{L}$ is defined by their quotient.
}
%Thus, a logical operator brings states within the code space to another state the code space, hence acts on the logical operators.
A logical operator operator is ``trivial'' if it is itself an element of $\mc{S}$, and ``nontrivial'' otherwise.
Consequently, a nontrivial logical operator acts \emph{within} the code space, but \emph{nontrivially}, and therefore is a so-called ``undetectable and uncorrectable error'' of the code.%\footnote{We will exclusively consider ``located errors'' in this paper, in which case all detectable errors can also be corrected.}

With this trivial/nontrivial distinction, it is clear that a logical operator is defined up to gauge freedom, that is, up to arbitrary multiplications of elements in $\mc{S}$ (which do not change its action on the code space).
Thus logical operators are most easily thought of as ``equivalence classes'', or formally, cosets of $\mc{S}$ in $\mc{C}(\mc{S})$.
We define the ``logical group'' $\mc{L}$ as the following quotient group, $\mc{L} \coloneqq \mc{C}(\mc{S}) / \mc{S}$,
with $|\mc{L}| = 2^{2k}$. We note that $\mc{L}$ can be generated by representative ``logical Pauli $X$- and $Z$-operators'', conventionally denoted as $\{\ovl{X}_{1\ldots k}, \ovl{Z}_{1\ldots k} \}$.
}

\subsection{The circuit evolution in the stabilizer formalism \label{sec:qecc_dynamics}}

We briefly describe the circuit dynamics in the stabilizer formalism, with the help of notions introduced above.
%We will also briefly introduce the phase diagram of the random Clifford circuit model in Fig.~\ref{fig:upcircuit}, where $p$ is the only tuning paramter of the model.
%Statements within this subsection hold generally for all stabilizer codes, unless otherwise noted.
We show that the state at any point of the circuit evolution, as governed by Eqs.~(\ref{eq:gate_U}, \ref{eq:gate_P}), remains a code state as in Eq.~\eqref{eq:rho_S}.
\env{enumerate}{
\item
Firstly, we notice that the initial maximally-mixed state is a code state with $\mc{S} = \emptyset$.
\item
Under a Clifford unitary gate $U$ as in Eq.~\eqref{eq:gate_U}
\env{align}{
    \rho_Q(\mc{S}) \to U \rho_Q(\mc{S}) U^\dg  = \rho_Q({U \mc{S} U^\dg}),
}
where $\mc{S}^\p = U \mc{S} U^\dg$ is obtained from $\mc{S}$ by conjugating each element of $\mc{S}$ by $U$.  Thus $\mc{S}^\p$ is also an abelian subgroup of $\mc{P}(Q)$.
Moreover, we have $S(\rho_Q(\mc{S})) = S(\rho_Q({\mc{S}^\p}))$  .
\item
Under a Pauli measurement of $g \in \mc{P}(Q)$ as in Eq.~\eqref{eq:gate_P}\footnote{The results here holds generally for all Pauli operators $g$, although we are mostly interested in single site Pauli operators that are relevant in the context of the circuit model in Fig.~\ref{fig:upcircuit} and in Eq.~\eqref{eq:gate_P}.}, one can easily verify that~\cite{aaronson0406chp}
\env{enumerate}{[(i)]
\item
When $g$ anticommutes with some elements of $\mc{S}$ (hence a detectable error), it is always possible to choose $\mc{G}$ such that it has \emph{exactly one} element that anticommutes with $g$.
The updated stabilizer group $\mc{S}^\p$ is generated by $\mc{G}^\p$, where 
\env{align}{
    \mc{G}^\p = \{g_j | g_j \in \mc{G}, [g_j, g] =0 \} \cup \{g \}.
}
\item
When $g$ commutes with all elements in $\mc{S}$ and is itself within $\mc{S}$ (a trivial logical operator/trivial error),
\env{align}{
    \mc{G}^\p = \mc{G}.
}
\item
When $g$ commutes with all elements in $\mc{S}$ and is itself \emph{not} within $\mc{S}$ (a nontrivial logical operator/undetectable error),
\env{align}{
    \mc{G}^\p = \mc{G} \cup \{g\}.
}
}
}
Thus, if the initial state of the circuit is a ``code state'' as in Eq.~\eqref{eq:rho_S}, then at any point of the circuit evolution the state is a code state, which admits an efficient representation in terms of $\mc{G}$, and \YL{consequently} efficient simulation of the circuit dynamics, \YL{a result known as the Gottesman-Knill theorem}~\cite{gottesman9807heisenberg, aaronson0406chp}.

Moreover, one sees from above that the entropy of the state decreases by $\ln 2$ (i.e. the state gets ``purified'' by one unit) if a nontrivial logical operator (or equivalently an ``undetectable error'') $g$ is measured (compare Eq.~\eqref{eq:s_rho_Q_G}), but remains unchanged otherwise~\cite{fan2020selforganized, ippoliti2004measurementonly}.
This observation provides a first clue to a possible connection between the error correcting properties of the state (when viewed as a QECC) and the purification dynamics.%, as we will elaborate next.

\subsection{Code distance and the theorem \label{sec:theorem}}

An important metric of a QECC is its ``code distance'', $d$, defined to be the minimal weight of all \YL{nontrivial} logical operators.\footnote{Recall that the weight of a Pauli string operator is the number of qubits on which its content is not the identity operator.}
In our circuit model that has locality, it is natural to define a similar quantity, the ``contiguous code distance''~\cite{gullans1905purification, ippoliti2004measurementonly}, $d_{\rm cont}$, as the minimal length of a contiguous segment of qubits that supports a \YL{nontrivial} logical operator.
\YL{By definition, $d \le d_{\rm cont}$.}

%Given a subset $A$ of qubits, the logical operators that can be supported on $A$ (that is, all elements of $\mc{C}(\mc{S})$ that have trivial contents on $\ovl{A}$) form a (non-abelian) subgroup of $\mc{C}(\mc{S})$, which we will call $\mc{L}(A)$.
%As a Pauli subgroup, we have $|\mc{L}(A)| = 2^{\ell_A}$, where $\ell_A$ is an integer, and has the meaning of ``the number of independent logical operators'' on $A$.
%The quantity $\ell_A$ thus measures how susceptible the QECC is to errors on $A$.

We say that a logical operator $g \in \mc{C}(\mc{S})$ is ``localizable'' on a set $A$ of qubits, if there exists $g^\p \in \mc{S}$ such that $g g^\p$ acts trivially on $\ovl{A}$, where $\ovl{A} \coloneqq Q - A$ is the complement of $A$ in $Q$.
It can be verified that all logical operators localizable on a given set $A$ form a subgroup of $\mc{C}(\mc{S})$.
It can also be verified that this subgroup \YL{of operators localizable on $A$} contains $\mc{S}$ as a subgroup, upon taking $g^\p = g \in \mc{S}$ above.
We take the quotient between these two, and denote the corresponding quotient group as $\mc{L}_A$, which is a subgroup of $\mc{L}$ (\YL{see Appendix~\ref{app:proof} for a detailed characterization of $\mc{L}_A$}).
%\env{align}{
    %\mc{L}_A \coloneqq \frac{\mc{C}(\mc{S}) \cap \(\mathrm{proj}_{\ovl{A}}\)^{-1} \lz \mathrm{proj}_{\ovl{A}}(\mc{S}) \rz} {\mc{S}},
%    \mc{L}_A \coloneqq \frac{\ld g \in \mc{C}(\mc{S}) \ |\  \mathrm{proj}_{\ovl{A}}(g) \in \mathrm{proj}_{\ovl{A}}(\mc{S}) \rd} {\mc{S}} \subseteq \mc{L}.
%}
We have $|\mc{L}_A| = 2^{\ell_A}$, where $\ell_A$ is an integer, and has the meaning of ``the maximal number of independent and inequivalent logical operators (undetectable errors)" on $A$.
The quantity $\ell_A$ thus measures how susceptible the QECC is to undetectable errors on $A$.

By definition, any subset (resp. segment) $A$ of qubits with weight (resp. length) smaller than $d$ (resp. $d_{\rm cont}$) supports no logical operators (therefore $\ell_A = 0$), or equivalently, no ``undetectable errors''.
An error occuring on $A$ must therefore be either ``detectable'' (that brings states outside the code space) or ``trivial'' (that leaves states within the code space unchanged).
When a detectable error located on $A$ occurs, an error correcting unitary supported on $A$ that reverses the effect of the error can be found, given its error syndrome~\cite{gottesman1997thesis, nielsen2010qiqc}.

Following the standard nomenclature, we may say that the circuit defines a $[|Q|, k, d_{\rm cont}]$-code over the course of its time evolution, where both $k$ and $d_{\rm cont}$ are functions of time.
A central purpose of this paper is to characterize the code dynamics, and develop an intuitive picture of its error correcting capabilities as quantified by $k$ and $d_{\rm cont}$.
This is %\st{partially}
{partly} achieved by the following relation between $\ell_A$ and the entanglement structure of the state:
%With these notions and motivations, we are now ready to state the main result of this section.

\textbf{Theorem 1.} Let $\rho_Q$ be a code state \YL{(defined in Eq.~\eqref{eq:rho_S} to be the maximally-mixed state on the code space)}, and $\ket{\Psi_{QR}}$ be a\YL{n arbitrary} purification of $\rho_Q$, %that is,
\env{align}{
    \rho_Q = {\rm Tr}_R \ket{\Psi_{QR}} \bra{\Psi_{QR}}.
}
Then for any subset $A$ of $Q$, $A \subseteq Q$, we have
\env{align}{
    \label{eq:ell_A_I_AR}
    \ell_A = (\ln 2)^{-1} I_{A, R},
}
where the RHS is the mutual information between $A$ and $R$, %i.e.
\env{align}{
    \label{eq:def_IAR}
    &\  I_{A, R} \nn
    =&\ S(\rho_A) + S(\rho_R) - S(\rho_{AR}) \nn
    =&\ S(\rho_A) + S(\rho_Q) - S(\rho_{\ovl{A}}).
}
Here again, $\ovl{A} \coloneqq Q - A$ is the complement of $A$ on $Q$.  {\hfill $\Box$}

The proof of the theorem is given in Appendix~\ref{app:proof}.

Several comments are in order:
\env{itemize}{
\item
The quantity $\ell_A$ was %first
introduced and explored in Refs.~\cite{yoshida1002classifying, haah1011logical} (see also Ref.~\cite{haah1607lecture}), although not explictly cast in the form of a mutual information.
From the theorem it follows directly that $\ell_A + \ell_{\ovl{A}} = 2k$, the ``cleaning lemma''~\cite{haah1607lecture, bravyi0810nogo}.
\item
Clearly, from its definition, $\ell_A \le \ell_{AB}$ \YL{since $A \subseteq AB$}, which implies $I_{A, R} \le I_{AB, R}$ or, equivalently, 
\env{align}{
    S(\rho_A) + S(\rho_{ABR}) \le S(\rho_{AB}) +  S(\rho_{AR}),
}
the strong subadditivity inequality.
\item
We have not specified the pure state $\ket{\Psi_{QR}}$.  However, since both sides of Eq.~\eqref{eq:ell_A_I_AR} can be defined from $\rho_Q$ alone (see the last line of Eq.~\eqref{eq:def_IAR}), %so that
any purification of $\rho_Q$ would work equally well.
\item
For concreteness, let us choose $|R| = k$, the minimal number of qubits required, and consider the following ``encoded state'' \YL{as a purification of $\rho_Q$},
\env{align}{
    \ket{\Psi_{QR}} = \frac{1}{\sqrt{2^k}} \sum_{x} \ket{\ovl{x}_Q} \ket{x_R},
}
where $\{ \ket{\ovl{x}_Q} \}$ is an orthonormal basis of the code space, and $\{ \ket{x_R} \}$ is an orthonormal basis of $R$.
This pure state can be obtained by starting from $k$ Bell pairs, \YL{collecting one qubit from each pair, and encoding this collection of $k$ qubits in the QECC (on $Q$)} while labelling the other $k$ qubits as $R$.

The implication of the theorem when \YL{$A$ is a contiguous segment and} $|A| < d_{\rm cont}$ is important, and perhaps familiar from general %\st{results}
\YL{considerations} of QECCs. %quantum error correcting codes.
The LHS of Eq.~\eqref{eq:ell_A_I_AR} is zero, following the definition of $d_{\rm cont}$.
The RHS is therefore also zero, as it must be~\cite{preskill1998lecture}:
Since all errors on $A$ can be detected (and hence corrected), no observables on $A$ can reveal any information about the encoded state, and there should be no correlations between $A$ and $R$.\footnote{In particular, no measurements on $A$ should be able to change the entropy of $\rho_Q(\mc{S})$.
The ``decoupling condition'' in Eq.~\eqref{eq:decouple} was argued to hold for typical states when $p < p_c$, thus responsible for the very exsitence of a ``mixed'' phase~\cite{choi2019qec}. We will come back to these points in Sec.~\ref{sec:discussion}.}
Therefore, they must ``decouple'' on the level of density matrices~\cite{schumacher_westmoreland_2001_approximateQEC, devetak2003private, hayden0702decoupling, winter2009mother},
\env{align}{
    \label{eq:decouple}
    |A| < d_{\rm cont} \quad \Rightarrow \quad
    \rho_{AR} = \rho_A \otimes \rho_R,
}
leading to %\st{zero}
\YL{a vanishing} mutual information between $A$ and $R$.
}
On a practical level, the theorem provides a concrete relationship between error correcting capabilities of the QECC and its entanglement structure.
For example, one can readily ``read off'' the code distance of the QECC, assuming a complete knowledge of the entanglement structure.

\section{Domain wall picture of entanglement entropies \label{sec:dw_picture}}

In this section we review the mapping of the circuit dynamics to effective stat. mech. models, as first developed for unitary Haar circuits in Refs.~\cite{Hayden2016, nahum2017KPZ, nahum2018operator, zhou1804emergent, zhou1912membrane}, and later extended to hybrid Haar circuits in Refs.~\cite{choi2019spin, andreas2019hybrid}.
In either case, the entanglement entropy can be related to a domain wall free energy in the stat. mech. model, which can receive both ``energetic'' and ``entropic'' contributions.
We will however focus on the case of hybrid circuits with a nonzero measurement strength $p$, where results in Refs.~\cite{andreas2019hybrid, choi2019spin} can be directly applied.

\subsection{Mapping to a spin model}

%The description of entanglement entropies in terms of ``domain walls'' is not a new one.
%It goes back to the Ryu-Takayanagi formula, and is also proposed in the context of random unitary circuits \emph{without} measurements~\cite{nahum2017KPZ, nahum2018operator, nahum1803coarsegrained, zhou1804emergent, zhou1912membrane}.

%----------------------------
\begin{figure}[t]
    \centering
    \includegraphics[width=.5\textwidth]{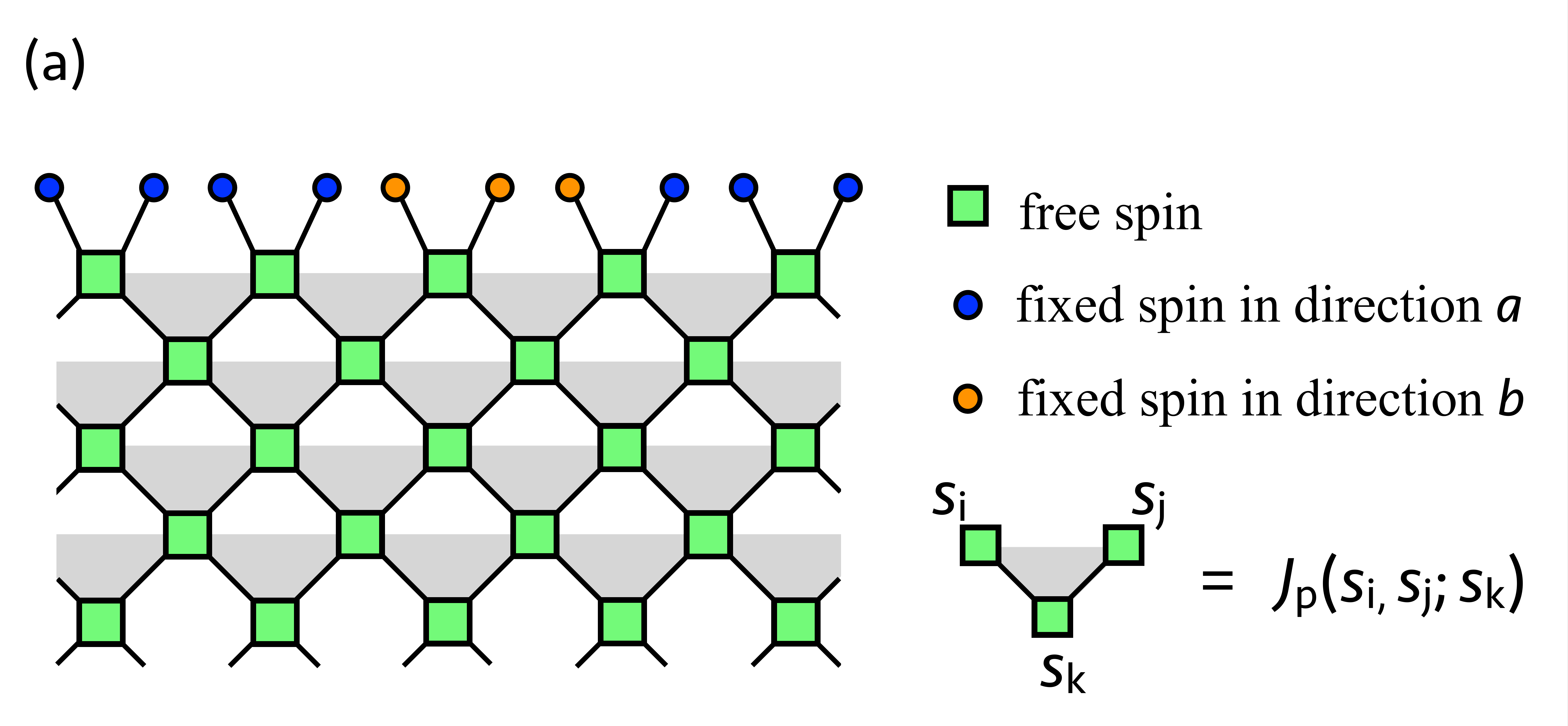}
    \includegraphics[width=.5\textwidth]{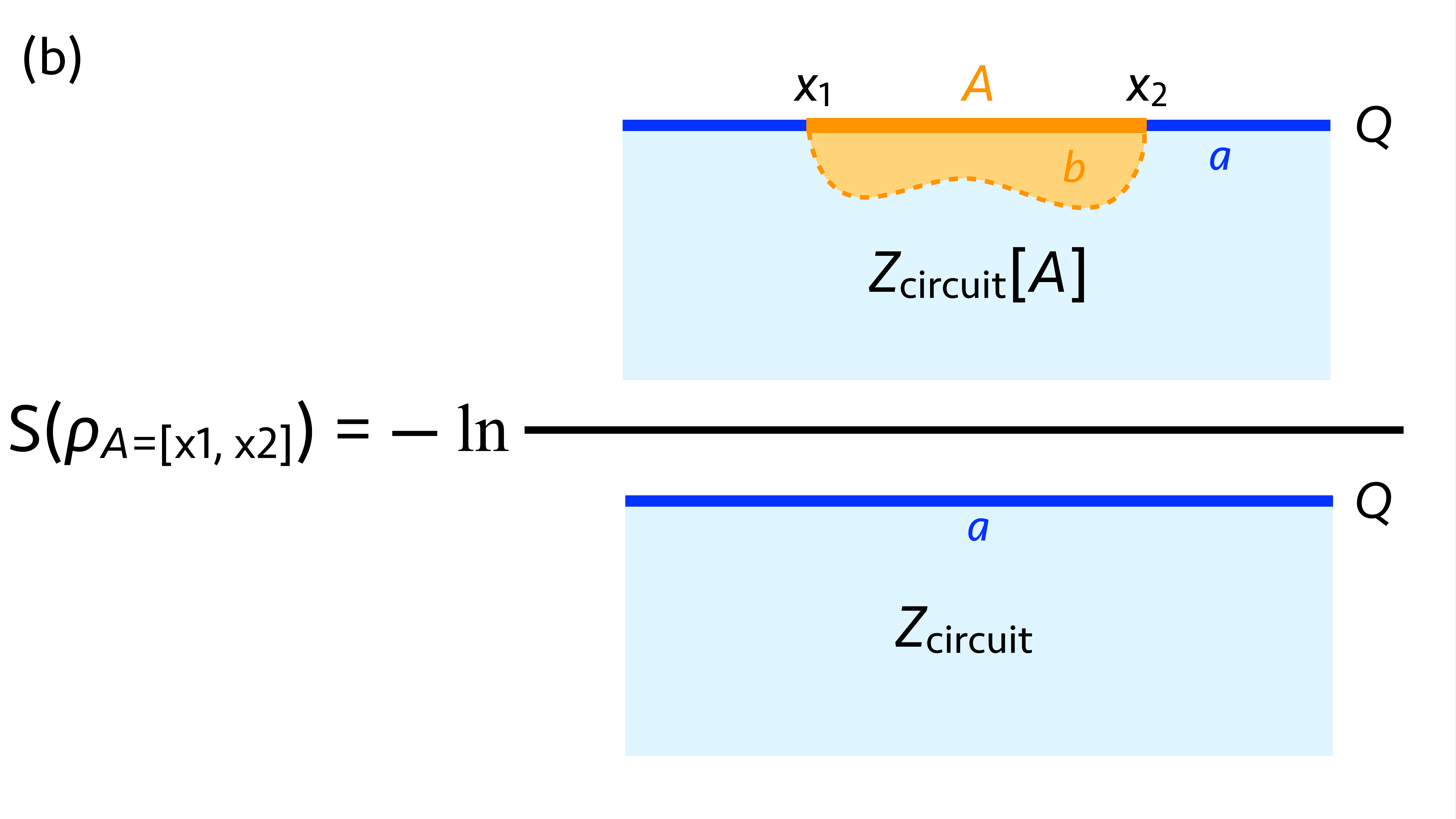}
\caption{
(a)
Illustration of the underlying spin model  in Refs.~\cite{choi2019spin, andreas2019hybrid} for the hybird random Haar circuit.
This figure is adapted from Ref.~\cite{andreas2019hybrid}.
Each bulk unitary maps to a bulk spin (green square), that is ``free''.
Qubits at the final time $t=T$ (solid dots) correspond to ``fixed'' spins all pointing in either the $\bc{a}$ or $\bc{b}$ direction.
A Boltzmann weight is associated with each downward-pointing triangle (shaded), and is a function of spins on its vertices.
(b)
Representation of the entanglement entropy of the segment $A$ as the difference of two free energies (see main text).
In this figure, we have chosen to ``zoom in'' on a small part near the upper edge of the circuit; the upper edge corresponds to physical qubits $Q$ at the final time $t=T$ of the circuit evolution.
The other boundaries are far away, and need not be specified.
The illustrations on the RHS represent typical configurations in the low-temperature ``ferromagnetic'' phase, possibly after a sufficient number of coarse-graining steps.
In the denominator ($Z_{\rm circuit}$), the upper edge is colored blue, corresponding to the fixed boundary condition $\bc{a}$; thus, the bulk spins tend to also order along $\bc{a}$.
In the numerator ($Z_{\rm circuit}[A]$) the segment $A$ is colored yellow, and the spins are aligned to have a different value $\bc{b}$.
This will induce the alignment of proximate bulk spins along the same direction $\bc{b}$.
A domain wall is then present where the two domains meet.
}
\label{fig:bc_ab}
\end{figure}
%----------------------------

The upshot of the mapping introduced in Refs.~\cite{choi2019spin, andreas2019hybrid} for the hybrid random Haar circuit
can be very roughly summarized as follows (compare Fig.~\ref{fig:bc_ab}(a)), where we omit technical details.
\YL{Recall that the hybrid random Haar circuit~\cite{choi2019spin, andreas2019hybrid} is structurally identical to the circuit in Fig.~\ref{fig:upcircuit}, except with each unitary gate sampled from the Haar measure on $\mathsf{U}(4)$, and the sporadic projective measurements replaced by generalized weak measurements of the same strength on each qubit at each time step.
This strength plays a role similar to the frequency of sporadic projective measurements in Fig.~\ref{fig:upcircuit}.
}

\env{enumerate}{
\item
In the bulk of the circuit,
there is one Potts-like spin degree-of-freedom associated with each unitary gate, taking values in $\{\bc{a}, \bc{b}, \bc{c}, \ldots \}$.
The bulk spins form a square lattice (see Fig.~\ref{fig:bc_ab}(a), \YL{and compare with Fig.~\ref{fig:upcircuit}}).
A Boltzmann weight is defined on each downward-pointing triangle of the lattice (see Fig.~\ref{fig:bc_ab}(a)).
The ``circuit partition function'', $Z_{\rm circuit}$, is obtained by contracting all free spin indices of the Boltzmann weights. %, with the latter viewed as tensors.
% By construction, when talking about the circuit partition function, one needs to specify the initial state and the spatial boundary conditions.

\item
At the $t = T$ (final time) boundary of the circuit, there is a spin associated with each physical qubit in $Q$.
All spins in $Q$ are fixed to have the same value (say $\bc{a}$), and therefore corresponds to a ``fixed'' boundary condition (b.c.) of the spin model.

\item
At the $t = 0$ (initial time) boundary of the circuit, there is also a spin associated with each physical qubit in $Q$, and certain short-range entangled initial states on $Q$ corresponds to simple b.c. of the spin model.
In particular, a pure product initial state corresponds to a ``free'' b.c., where each spin can indepedently take all allowed values.  On the other hand,
the maximally-mixed initial state corresponds to the same ``fixed'' b.c. ($\bc{a}$) as at the $t=T$ boundary.

\item
When the spatial b.c. is periodic, the circuit geometry is cylindrical, and there are no other boundaries of the circuit.
When the spatial b.c. is open, the circuit geometry is rectangular, and the boundary conditions on the left and right sides of the rectangle are also ``free''.

\item
To compute the entanglement entropy of a segment $A=[x_1, x_2]$ in the final state (at $t=T$), one needs to compute another partition function.
This partition function is defined by the same boundary condition as $Z_{\rm circuit}$, except with all spins in the segment $A$ at the upper edge of the circuit ``aligned'' to another different value, say $\bc{b}$.
%Define complex coordinates $z_{1,2} = x_{1,2} + iT$, 
We call this new partition function $Z_{\rm circuit}[A]$. %, where $\phi(z_{1,2})$ are ``boundary condition changing (b.c.c.)'' operators signifying the change of b.c. at endpoints of $A$.

As shown in~\cite{andreas2019hybrid, choi2019spin, vasseur2018rtn}, the (ensemble averaged) entanglement entropy follows,
\env{align}{
\label{eq:SA_log_Z}
    S(\rho_{A}) = -\ln \frac{Z_{\rm circuit}[A]}{Z_{\rm circuit}},
}
 taking the form of a ``free energy cost'' due to the change of b.c. (see Fig.~\ref{fig:bc_ab}(b)).

\item
The purification/entanglement transition corresponds to an ordering transition of the spin model, where the measurement strength $p$ plays a role similar to temperature.
Within the low-temperature ordered phase of the spin model, a well-defined domain wall with finite surface tension must be present to account for the b.c. change (see Fig.~\ref{fig:bc_ab}(b)).\footnote{Notice that in Fig.~\ref{fig:bc_ab}(b), we have chosen to present the domain wall in the simplest form, where $\bc{a}$ and $\bc{b}$ can meet directly (so that there can be as few as only one domain wall) and are the only spin values that need to be considered.
The spins are therefore Ising like.
In general, it might be energetically favorable to have yet other different domains inserted between the $\bc{a}$- and $\bc{b}$-domains in typical configurations subject to this b.c., resulting in multiple mutually-avoiding domain walls~\cite{andreas2019hybrid}.
}
The free energy cost, mostly coming from the domain wall, will be extensive, leading to volume law entanglement entropies.
We will, for brevity, call it ``the entanglement domain wall''.
The surface tension decreases with increasing $p$, and eventually vanishes at the critical point.
}

As demonstrated for random Haar circuits~\cite{choi2019spin, andreas2019hybrid}, this mapping requires a replica limit of the spin model (the limit where the \YL{number of} available values of the Potts spins goes to $1$), and enables certain predictions for critical properties of the model~\cite{andreas2019hybrid} for $n\ge 1$ R\'{e}nyi entropies.
However, the more general viewpoint of entanglement entropies (namely as free energies of domain walls)~\cite{nahum2017KPZ, nahum2018operator, zhou1804emergent, zhou1912membrane, nahum2017quenched, nahum1803coarsegrained} has proven useful in understanding the phase transition in other contexts: for the zeroth R\'{e}nyi entropy~\cite{nahum2018hybrid} (where the entanglement entropy is equal to a ``geometrical minimal cut'' of the underlying lattice); and for critical properties of the random Clifford circuit~\cite{gullans1910scalable, li2003cft}.

We will henceforth assume this general domain wall picture holds for the random Clifford circuit in the mixed phase. % in both phases, as well as the critical point.
Since a derivation of the underlying stat.~mech. model (if it exists) is unavailable at present,
the precise nature of the domain walls is unknown.
Nevertheless, as we shall see, the domain wall picture alone, with the additional assumption that the domain walls are of the simplest type (``Ising like''; see Fig.~\ref{fig:bc_ab}), captures much of the qualitative aspects of the entanglement entropies in the Clifford circuit.
We will devote the rest of this section to capillary-wave theory of Ising domain walls and its implications, and the next section to numerical checks of capillary-wave theory for the Clifford circuit.

\subsection{Capillary-wave theory of Ising domain walls}

Capillary-wave theory~\cite{buff1965capillary, weeks1977capillary, mpaf1982capillary} was originally proposed for describing domain walls in the low-temperature ordered phase of the Ising model.
For the example in Fig.~\ref{fig:bc_ab}, a sharp domain wall must be present to be consistent with the assigned boundary conditions.
For this geometry, one can further argue that it is sufficient to consider configurations with a single domain wall, which also admits the following parametrization as a ``height function'',
\env{align}{
    \label{eq:height_fcn}
    y : [x_1, x_2] \to&\ [-T, 0], \nn
        x \mapsto&\ y(x),
    %y(x), \text{where } x \in [x_1, x_2] \text{ and } y(x_1) = y(x_2) = 0.
}
where $y(x_1) = y(x_2) = 0$.
%The domain wall is then a ``directed walk'', with $x$ playing the role of ``time'' and $y$ playing the role of ``space''.
With this parameterization, we are %\st{ignoring}
\YL{neglecting} all ``overhangs'' and ``bubbles'' that might be present in the relevant configurations; these have a finite typical size in the low-temperature phase, and will eventually disappear under coarse-graining.
This reasoning leads to the following approximation for the entanglement entropy,
\env{align}{
    \label{eq:Z_y_x_Gaussian}
    &  S(\rho_{A = [x_1, x_2]}) \nn
    =& -\ln \frac{Z_{\rm circuit}[A]}{Z_{\rm circuit}} \nn
    %\approx& -\ln \frac{Z_{\rm circuit} \int \mc{D}[y(x)] \exp \lz -\beta \sigma \int_{x_1}^{x_2} dx \sqrt{1 + \( \pd_x y \)^2} \rz}{Z_{\rm circuit}} \nn
    \approx& -\ln \int \mc{D}[y(x)] \exp \lz -\beta \sigma \int_{x_1}^{x_2} dx \sqrt{1 + \( \pd_x y \)^2} \rz,
}
where $\beta$ is the ``inverse temperature'', and $\sigma$ the ``surface tension''.
This resulting ``capillary-wave theory'' partition function is the canonical ensemble of all domain walls (i.e. height functions $y(x)$ defined in Eq.~\eqref{eq:height_fcn}), where the energy of each domain wall is the product of the surface tension and its surface area.
%Thus, within this approximation, the free energy cost is approximately by the domain wall.

After expanding the square root and dropping higher-order irrelevant terms, Eq.~\eqref{eq:Z_y_x_Gaussian} becomes a Gaussian theory, and can be readily evaluated.  With details in Appendix~\ref{app:cw}, we find,
\env{align}{
    \label{eq:s_rho_A_small}
    &\  S(\rho_A) \nn 
    \approx&\ F_\mathrm{CW}(A) \nn%, \text{ when } |A| \ll L \nn
    =&\ \beta \sigma |A| + \frac{3}{2} \ln |A|, \text{ when } T \gg \sqrt{L} \gg \sqrt{|A|},
}
for $|A| \gg 1$.
Here the first term is the surface energy, and the second term is ``entropic'', coming from transverse, thermal fluctuations of $y(x)$, with a universal coefficient $3/2$, as found in Ref.~\cite{fan2020selforganized} within a quantized regularization of the Ising model. % Eq.~\eqref{eq:Z_y_x_Gaussian}.
Notice that we have reserved the notation $S(\rho_A)$ for the entanglement entropy of $A$, and $F_\mathrm{CW}(A)$ for the free energy of the domain wall due to a change of b.c. in $A$.

In Fig.~\ref{fig:bc_ab} we have not specified boundary conditions on the lower-, left-, and right-sides of the circuit, as it is a ``zoomed-in'' view.
In this way, we are assuming implicitly that $|A| \ll L$, and also that the circuit depth is large compared to the vertical extent of the domain wall, $T \gg \sqrt{|A|}$.
It is within this regime that the approximation of $S(\rho_A)$ with $F_\mathrm{CW}(A)$ in Eq.~\eqref{eq:s_rho_A_small} is established, and is valid regardless of the other boundary conditions (as the domain wall is sufficiently far away from the other boundaries).
%This approximation is no longer valid when $|A|$ approaches $L$.

The subleading ``entropic'' correction of $S(\rho_A)$ was found to be characteristic of the mixed phase $0 < p < p_c$ of hybrid circuits~\cite{li1901hybrid, gullans1905purification}, and is now shown to be present generically whenever the fluctuating domain wall picture is valid,
though its analytic form ($\frac{3}{2} \ln |A|$) here is special to capillary-wave theory.
Its importance will be made clear in the next subsection.

We conclude this subsection by mentioning the limit $p = 0$, which corresponds to a random unitary circuit \emph{without} measurements.
In this case, the entanglement domain walls are directed in the temporal direction of the circuit (as opposed to the case here with $p > 0$, in Fig.~\ref{fig:bc_ab}, where the domain wall is directed in the spatial direction of the circuit).
This domain wall can now fluctuate in the transverse (spatial) direction, and these fluctuations leads to a similar entropic term $\frac{1}{2} \ln t$ when $t \ll L$~\cite{zhou1804emergent}, where the coefficient $\frac{1}{2}$ is universal, and also comes from the diffusion equation (see Appendix~\ref{app:cw}).
However, this term will disappear as $t = \mathrm{poly}(L) \gg L$, the regime we focus on in this paper.\footnote{\label{fn:adam_tianci}We thank Tianci Zhou and Adam Nahum for explaining to us Ref.~\cite{zhou1804emergent} on these points.}

\subsection{The maximally-mixed initial state}

%----------------------------
\begin{figure}[t]
    \centering
    \includegraphics[width=.5\textwidth]{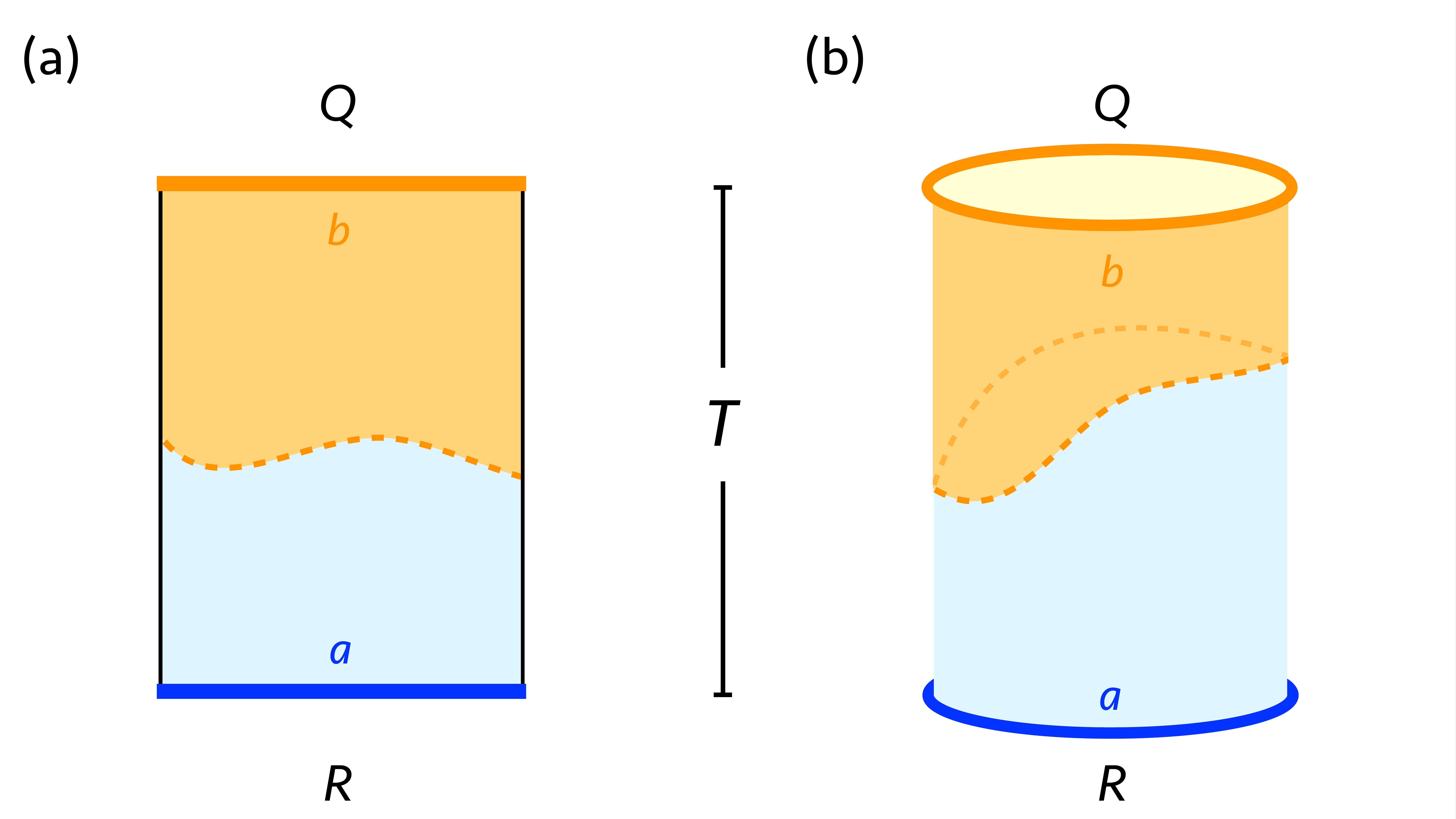}
\caption{
Illustrations of boundary conditions for $F_\mathrm{CW}(Q)$ with open b.c. (left) and periodic b.c. (right).
It is understood that both $F_\mathrm{CW}(Q)$ are obtained by subtracting the background free energy with both $Q$ and $R$ fixed to have b.c. $\bc{a}$ (not plotted), from the free energy of the configuration plotted (with $Q$ in $\bc{b}$ and $R$ in $\bc{a}$).
}
\label{fig:bc_waist}
\end{figure}
%----------------------------

We have seen in Sec.~\ref{sec:qecc_dynamics} that the hybrid circuit dynamics with the maximally-mixed initial state can be formulated as the dynamics of the correpsonding QECC.
The entropy of the entire system $Q$, $S(\rho_Q)$, is monotonically decreasing, corresponding to a monotonically decreasing ``code rate''.
On the other hand,
according to the prescriptions summarized at the beginning of this section (for mapping to a spin model),
the corresponding circuit partition function $Z_{\rm circuit}$ is defined by the fixed b.c. $\bc{a}$ on both the upper and lower edges of the circuit~\cite{choi2019spin, li2003cft}; whereas the entropy $S(\rho_Q)$ is the change in free energy upon changing the b.c. of the upper edge (i.e. qubits in $Q$) to a different, fixed one, $\bc{b}$.
The b.c. relevant to this calculation is illustrated in Fig.~\ref{fig:bc_waist}.
Since the maximally-mixed initial state admits a natural purification in terms of  $|Q| = L$ Bell pairs where $Q$ consists of one qubit from each pair, the upper and lower edges can be naturally viewed as $Q$ (the system, that is acted upon by the circuit), and $R$ (the ``reference'', consisting of the other half of the Bell pairs, that is left un-evolved by the circuit), respectively~\cite{li2003cft}.
We will henceforth adopt this labelling, for we find it intuitive to have a concrete reference $R$ at the far end of the circuit that $Q$ is trying to disentangle itself from, even if this choice of $R$ is not unique.

The dominant contribution to $S(\rho_Q)$ comes from a single domain wall separating the upper and lower edges, going around the ``waist'' of the circuit (again compare Fig.~\ref{fig:bc_waist}):
\env{itemize}{
\item
With open spatial b.c., the domain wall endpoints are ``free'', and can independently take any vertical coordinate $y(x=0)\in [-T, 0]$ and $y(x=L) \in [-T, 0]$.
\item
With periodic spatial b.c., the domain wall is periodic, but otherwise ``free'' to take any position along the vertical direction, leading to $y(x=0) = y(x=L) \in [-T, 0]$.
}
The free energies can then be calculated within capillary-wave theory (Appendix~\ref{app:cw}),
\env{align}{
    \label{eq:s_rho_Q}
    &\ S(\rho_Q) \nn
    =& -\ln \frac{Z_{\rm circuit}[Q]}{Z_{\rm circuit}} \nn
    \approx&\  F_{\rm CW}(Q) \nn
    =& \env{cases}{
        \beta \sigma L - \ln T, & \text{open b.c.}\\
        \beta \sigma L - \ln \frac{T}{\sqrt{L}},& \text{periodic b.c.}
    }
    \text{ when } T \gg \sqrt{L}.
}
The $-\ln T$ term comes from the ``center of mass entropy'' of the ``waist domain wall'', whose form is consistent with an exponentially long purification time within the mixed phase~\cite{gullans1905purification} (see Sec.~\ref{sec:crossover}). %; the long time behavior as $T \gg \exp(\beta \sigma L)$ can be captured by considering more than one non-interacting waist domain walls (see Appendix~\ref{app:cw}).
The $\ln\sqrt{L}$ difference between open and periodic b.c. is attributed to the additional endpoint entropy with open b.c., as mentioned above.

We see also that the quantity $\beta \sigma$ can be identified as ($\ln 2$ times) the code rate,
\env{align}{
    \label{eq:code_rate_surface_tension}
    \lim_{|Q| \to \infty} \frac{k\ln 2}{|Q|} = \lim_{L \to \infty} \frac{S(\rho_Q)}{L} = \beta \sigma, 
}
for $T = \textrm{poly}(L)$.

\subsection{Decoupling of domain walls}

We are now ready to investigate the entropy of a \emph{contiguous} subregion $A$ of $Q$ with arbitrary length.
Notice that the previous result in Eq.~\eqref{eq:s_rho_A_small} was obtained for $|A| \ll |Q| = L$, and that Eq.~\eqref{eq:s_rho_Q} accounts for the limiting case $|A| = |Q|= L$.
These two regimes must then be interpolated by some intermediate behavior.
For convenience, below, we will instead study the entropy $S(\rho_{\ovl{A}})$,
defined on the complement of $A$.  

Consider first the limit with small $A$, $|A| \ll |Q|$.
%to be small, and consider the entropy of its complement, $S(\rho_{\ovl{A}})$, while taking $|A| %\to 0$ (hence $|\ovl{A}| \to L$).
In this regime, the partition function $Z_{\rm circuit}[\ovl{A}]$ (defined by $\bc{a}$ on $AR$ and $\bc{b}$ on $\ovl{A}$) receives two possibly comparable contributions (see Fig.~\ref{fig:bc_decouple} with periodic b.c.):

%----------------------------
\begin{figure}[t]
    \centering
    \includegraphics[width=.5\textwidth]{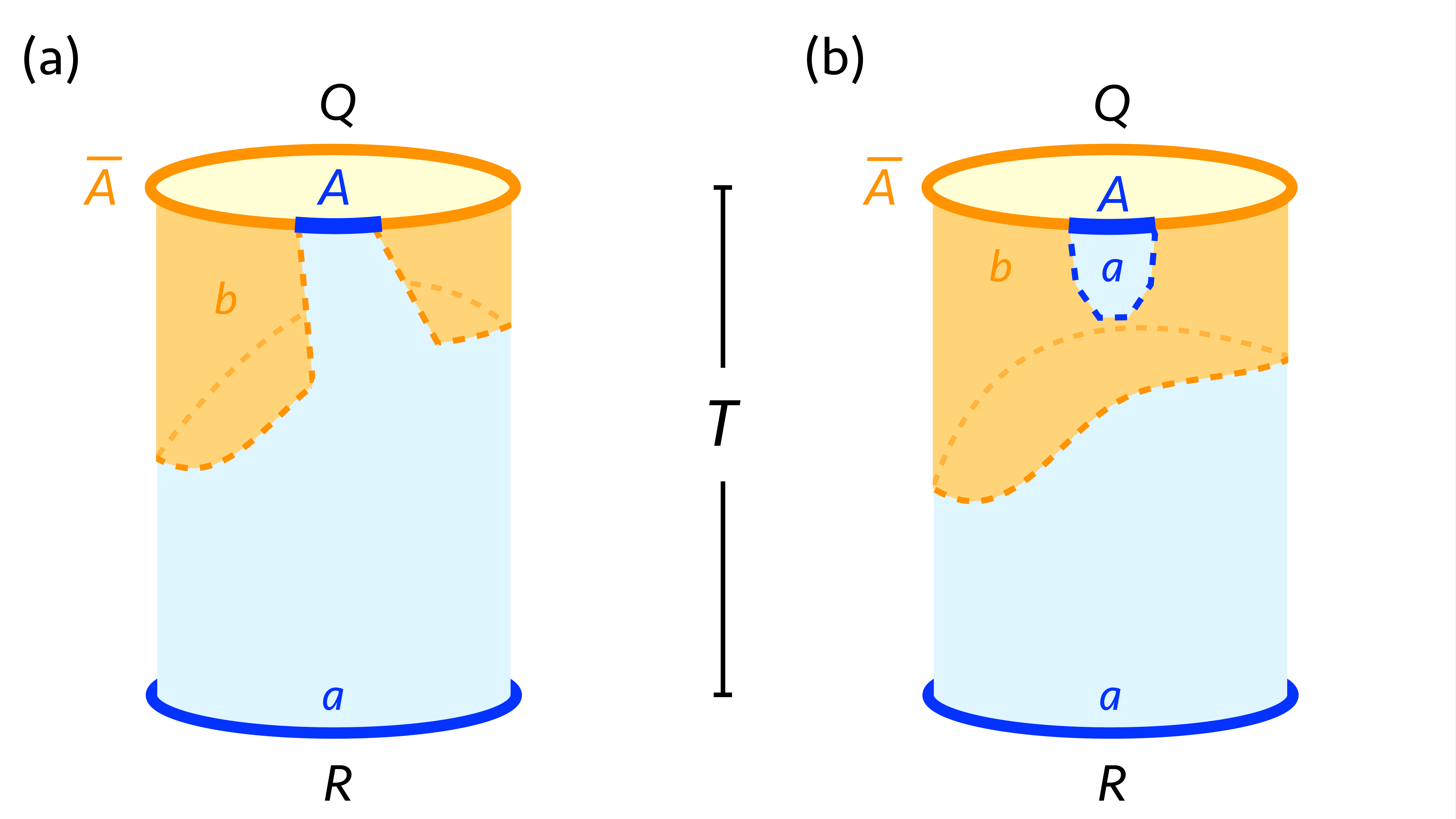}
    \includegraphics[width=.5\textwidth]{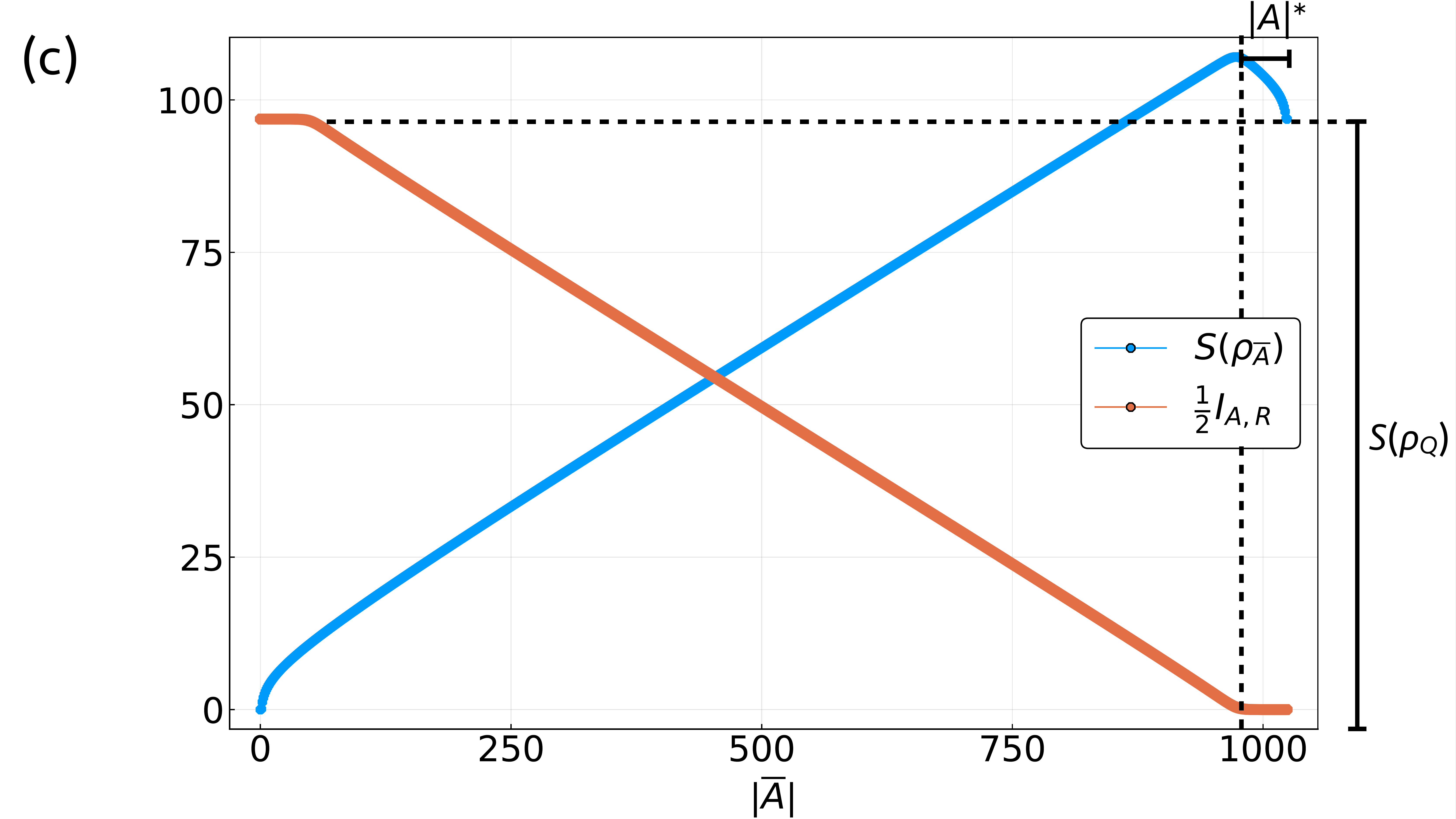}
\caption{
(a,b) Illustrations of boundary conditions for $F_\mathrm{CW}(\ovl{A})$ with periodic spatial boundary conditions.
The partition function $Z_{\rm circuit}[\ovl{A}]$ is the sum of the two contributions, $Z_{\rm circuit}[\ovl{A}] = Z_{\rm circuit}^{(1)}[\ovl{A}] + Z_{\rm circuit}^{(2)}[\ovl{A}]$.
(c) The resulting entanglement entropy $S(\rho_{\ovl{A}})$ and (half) the mutual information $I_{A,R}$, as computed from capillary-wave theory (specifically Eqs.~(\ref{eq:s_rho_A_small}, \ref{eq:s_rho_Q}, \ref{eq:s_A_f1_f2})).
We have taken $|Q| = L = 1024$, $T  = 8L$, and $\beta \sigma = 0.1$ in this plot.
We emphasize the non-monotonicity in $S(\rho_{\ovl{A}})$ as $|A| \to L$.
Moreover, there is a linearly decreasing segment of the $\frac{1}{2} I_{A,R}$ versus $|\ovl{A}|$ plot, with horizontal extent $L - 2|A|^\ast$ and vertical extent $S(\rho_Q)$.
Since its slope must be bounded between $[-\ln 2, \ln 2]$, we have $(\ln 2)^{-1} S(\rho_Q) \le L - 2|A|^\ast$.
}
\label{fig:bc_decouple}
\end{figure}
%----------------------------

\env{enumerate}{
\item
A single domain wall separating $\ovl{A}$ from $AR$ as before. There are then two domains, with spins aligned along $\bc{a}$ and $\bc{b}$, respectively (see Fig.~\ref{fig:bc_decouple}(a)).
The corresponding partition function is approximated within capillary wave theory as
\env{align}{
    Z_{\rm circuit}^{(1)}[\ovl{A}] \approx Z_{\rm circuit} \, e^{-F_{\rm CW}(\ovl{A})}.
}
\item
Two ``decoupled'' domain walls, one separating $A$ from $\ovl{A}$, and the other, a ``waist domain wall'', separating $Q = A \ovl{A}$ from $R$.
There are now three domains, as shown in Fig.~\ref{fig:bc_decouple}(b), and to go from $A$ to $R$ \emph{two domain walls must be crossed}.
The corresponding partition function is approximated within capillary wave theory as
\env{align}{
    Z_{\rm circuit}^{(2)}[\ovl{A}] \approx Z_{\rm circuit} \, e^{-F_{\rm CW}(A) - F_{\rm CW}(Q)}.
}
}
After summing these contributions, we have, according to Eq.~\eqref{eq:SA_log_Z},
\env{align}{
    \label{eq:s_A_f1_f2}
    S(\rho_{\ovl{A}})
    %=& -\ln \frac{Z_{\rm circuit}[\ovl{A}]}{Z_{\rm circuit}} \nn
    %\approx& -\ln \frac{1}{Z_{\rm circuit}}\( Z_{\rm circuit}^{(1)}[\ovl{A}] + Z_{\rm circuit}^{(2)}[\ovl{A}] \) \nn
    \approx -\ln \lz e^{-F_{\rm CW}(\ovl{A})} + e^{-F_{\rm CW}(A) - F_{\rm CW}(Q)} \rz. % \nn
    %\approx&\ \mathrm{min} \ld F_{\rm CW}(\ovl{A}), F_{\rm CW}(A) + F_{\rm CW}(Q) \rd. % \nn
    %\approx& \env{cases}{
    %    F_{\rm CW}(\ovl{A}),& 0 \le |\ovl{A}| \le L - |A|^\ast \\
    %    F_{\rm CW}(A) + F_{\rm CW}(Q),& L - |A|^\ast \le |\ovl{A}|  \le L
    %}
}
The first contribution $F^{(1)}_{\rm CW} = F_{\rm CW}(\ovl{A})$ is always \emph{energetically} more favorable than $F^{(2)}_{\rm CW} = F_{\rm CW}(A) + F_{\rm CW}(Q)$, but is not necessarily \emph{entropically} so.
The competition is only present due to fluctuations of the domain walls.\footnote{A similar competition between domain wall topologies is also present in the limit $p=0$~\cite{zhou1804emergent}, which leads to an $O(1)$ ``Page correction'' to the entanglement entropy.}

To illustrate this, we evaluate Eq.~\eqref{eq:s_A_f1_f2} with periodic b.c.
(using Eqs.~(\ref{eq:s_rho_A_small}, \ref{eq:s_rho_Q})), where $F_{\rm CW}$ is simply a function of the size of the region, and plot the result in Fig.~\ref{fig:bc_decouple}(c).
Notice the striking non-monotonic behavior in $S(\rho_{\ovl{A}})$, which has a width labelled as $|A|^\ast$. The non-monotonicity
comes from a competition between the two contributions, which we can readily %\st{understood}
\YL{understand} for large $|Q|=L$,
\env{align}{
    \label{eq:s_A_min_f1_f2}
    & S(\rho_{\ovl{A}}) \nn
    \approx& -\ln \lz e^{-F_{\rm CW}(\ovl{A})} + e^{-F_{\rm CW}(A) - F_{\rm CW}(Q)} \rz \nn
    \approx&\ \mathrm{min} \ld F_{\rm CW}(\ovl{A}), F_{\rm CW}(A) + F_{\rm CW}(Q) \rd \nn
    \approx& \env{cases}{
        F_{\rm CW}(\ovl{A}), & 0 \le |\ovl{A}| < L - |A|^\ast \\
        F_{\rm CW}(A) + F_{\rm CW}(Q), & L - |A|^\ast < |\ovl{A}| \le L  .
    }
}
Here $|A|^\ast$ is the length scale when the entropic and energetic terms are comparable, and may be defined as follows,
\env{align}{
    \label{eq:Astar_vs_L}
    & F_{\rm CW}(L - |A|^\ast) = F_{\rm CW}(|A|^\ast) + F_{\rm CW}(L) \nn
    \Rightarrow&\hspace{.66in} |A|^\ast \approx \frac{1}{2\beta \sigma} \(\frac{3}{2} \ln L + \ln \frac{T}{\sqrt{L}} \),
}
to leading order for large $T$ and $L$.
The length scale $|A|^\ast$ is thus inversely proportional to the code rate $\beta \sigma$, and grows with both $L$ and $T$ logarithmically.
%It eventually becomes infinite for a pure state, either at late times when $T$ is exponentially large in $L$; or as $p \to p_c$ where $\beta \sigma \to 0$.
For any circuit depth $T = O(\mathrm{poly}(L))$, $|A|^\ast$ is proportional to $\ln L$.

In the regime with
$L - |A|^\ast < |\ovl{A}| \le L$ (i.e. $0 \le |A| < |A|^\ast$), we recognize that the free energies $F_{\rm CW}(A)$ and $F_{\rm CW}(Q)$ in Eq.~\eqref{eq:s_A_min_f1_f2} represent the corresponding entanglement entropies $S(\rho_A)$ and $S(\rho_Q)$ according to Eqs.~(\ref{eq:s_rho_A_small}, \ref{eq:s_rho_Q}).
The last line in Eq.~\eqref{eq:s_A_min_f1_f2} can then be rewritten as,
\env{align}{
    \label{eq:decoupling_condition}
    %\YL{&\quad 0 \le \frac{|A|}{|A|^\ast} \le 1 \nn}
    \YL{0 \le \frac{|A|}{|A|^\ast} < 1}
    \Rightarrow&\quad
    %& \quad 
    S(\rho_{\ovl{A}}) \approx S(\rho_A) + S(\rho_Q) %+ O(e^{F_{\rm CW}(A) + F_{\rm CW}(Q) - F_{\rm CW}(\ovl{A})})
    \nn
    \Leftrightarrow&\quad S(\rho_{AR}) \approx S(\rho_A) + S(\rho_R) \nn %+ O(e^{F_{\rm CW}(A) + F_{\rm CW}(Q) - F_{\rm CW}(\ovl{A})}) \nn
    \Leftrightarrow&\quad I_{A,R} \approx 0.
}
We thereby conclude that if $\frac{|A|}{|A|^\ast} < 1$, the subsystems $A$ and $R$ decouple.
This decoupling corresponds to the regime 
where the configuration in Fig.~\ref{fig:bc_decouple}(b) dominates, i.e. when the domain wall decouples, with two domain walls separating $A$ and $R$.  

In Fig.~\ref{fig:bc_decouple}(c) we have also plotted \YL{(half)} the mutual information between $A$ and $R$,
$I_{A,R}$, as computed from capillary-wave theory using  Eqs.~(\ref{eq:def_IAR}, \ref{eq:s_rho_A_small}, \ref{eq:s_A_f1_f2}).
Notice the (near) vanishing of $I_{A,R}$ for $0 \le |A| < |A|^\ast$, consistent with 
Eq.~\eqref{eq:decoupling_condition}.

A more detailed calculation shows that, 
\env{align}{
    \label{eq:decoupling_condition_estimate}
    &\ I_{A, R} \nn
    \approx&\ \ln \lz 1 +  e^{F_{\rm CW}(A) + F_{\rm CW}(Q) - F_{\rm CW}(\ovl{A})} \rz \nn
    \approx&\ \ln \lz 1 + e^{- 2 \beta \sigma (|A|^\ast - |A|)} \rz \nn
    \approx&\ \env{cases}{
        e^{-2\beta \sigma (|A|^\ast - |A|)}, & 0 \le |A| < |A|^\ast; \\
        2 \beta \sigma (|A| - |A|^\ast), & |A| > |A|^\ast  .
    }. %\text{ where } x = \frac{|A|}{|A|^\ast}.
}
Here we have used Eq.~\eqref{eq:Astar_vs_L}, and only kept the leading linear terms in $F_{\rm CW}$.
Since $|A|^\ast$ diverges in the thermodynamic limit (see Eq.~\eqref{eq:Astar_vs_L}), for $\frac{|A|}{|A|^\ast} \in [0, 1)$
the mutual information $I_{A,R}$ vanishes \emph{exactly}.  On the other hand, $I_{A,R}$ is \emph{strictly} positive if $\frac{|A|}{|A|^\ast} > 1$.

Upon combining with Theorem 1 in Sec.~\ref{sec:theorem}, we conclude that $\ell_A = 0$ \emph{if and only if} $|A| < |A|^\ast$.  We can then make the important identification between the code distance $d_{\rm cont}$ and $|A|^\ast$,
\env{align}{
    d_{\rm cont} = |A|^\ast.
}
With this equality, we may deduce from Fig.~\ref{fig:bc_decouple}(c) (see the figure caption) that
\env{align}{
    k = (\ln 2)^{-1} S(\rho_Q) \le L - 2|A|^\ast = |Q| - 2d_{\rm cont}.
}
This is essentially the quantum Singleton bound~\cite{knill_laflamme_1997}, with $d \to d_{\rm cont} \gg 1$.

To summarize, capillary-wave theory predicts that the dynamically generated QECC has code distance that diverges with system size on relevant time scales, while also keeping a finite code rate.
Qubit segments with length smaller than $|A|^\ast = d_{\rm cont}$ are protected from undetectable errors by thermodynamic fluctuations of the entanglement domain walls.

\subsection{Crossover to late times \label{sec:crossover}}

In this subsection, we deviate from our main focus on polynomial time scales $T = O(\mathrm{poly}(L))$, and briefly discuss how the domain wall picture can account for the late time crossover behavior when $T$ is exponential in $L$.
On these time scales, the entropy of the code state $S(\rho_Q)$ is expected to decay to zero~\cite{gullans1905purification}, \YL{i.e. the state is completely purified.}

Previously, when computing $S(\rho_Q)$ from Fig.~\ref{fig:bc_waist} obtaining the result in Eq.~\eqref{eq:s_rho_Q}, we only took into account configurations with a single waist domain wall -- valid since the energy term $\beta \sigma L$ is always dominant over the entropy term $-\ln T$ when $T = O(\mathrm{poly}(L))$, and single-domain wall %\st{terms}
\YL{configurations} have the lowest energy.
This simplification eventually breaks down when $T \gg \exp\lz \beta \sigma L\rz$, and we have to consider the possibility of multiple waist domain walls.
In particular, $Z_{\rm circuit}$ will now receive contributions from all configuration with an \emph{even} number of waist domain walls, and $Z_{\rm circuit}[Q]$ from those with an \emph{odd} number.
Here we are again assuming the Ising nature of these domain walls.
Moreover, the vertical (i.e. time direction) extent of each domain wall scales as $\sqrt{L}$ (see Appendix~\ref{app:cw}), much smaller than either $L$ or $T$.
These domain walls are therefore effectively ``local'' along the time direction, and the only interaction between the domain walls is onsite repulsion (i.e. the domain walls cannot overlap/cross, but otherwise non-interacting).
We thus have a picture of a ``(waist) domain wall gas'',
and can readily compute the corresponding partition functions using Eq.~\eqref{eq:s_rho_Q},
\env{align}{
    &\ S(\rho_Q) \nn
    =& -\ln \frac{Z_{\rm circuit}[Q]}{Z_{\rm circuit}} \nn
    \approx& -\ln \frac{\sum_{n \text{ odd}} \, (1/n!) \exp \lz - n F_{\rm CW}(Q) \rz}{\sum_{n \text{ even}} (1/n!) \exp \lz - n F_{\rm CW}(Q) \rz} \nn
    \approx& -\ln \tanh \(e^{- F_{\rm CW}(Q)}\) \nn
    \approx& \env{cases}{
    -\ln \tanh \( T e^{-\beta \sigma L} \) &\text{open b.c.} \\
    -\ln \tanh \( \frac{T}{\sqrt{L}} e^{-\beta \sigma L} \) &\text{periodic b.c.} \\
    }.
}
Notice that
\env{align}{
    \lim_{T \to \infty} S(\rho_Q) = 0,
}
as expected for a pure state.

The same reasoning leads to a similar modification of $S(\rho_{\ovl{A}})$ in Eq.~\eqref{eq:s_A_f1_f2},
\env{align}{
    S(\rho_{\ovl{A}})
    \approx - \ln \lz e^{-F_{\rm CW}(\ovl{A})} + e^{-F_{\rm CW}(A)} \tanh \( e^{- F_{\rm CW}(Q)} \) \rz,
}
which implies,
\env{align}{
    & \lim_{T \to \infty} S(\rho_{\ovl{A}}) \nn
    \approx& 
    - \ln \lz e^{-F_{\rm CW}(\ovl{A})} + e^{-F_{\rm CW}(A)} \rz \nn
    \approx&\ \mathrm{min} \ld F_{\rm CW}(\ovl{A}), F_{\rm CW}(A) \rd,
}
again as expected for the pure state $\rho_Q$ that is dynamically generated on exponentially long times.

Inclusion of multiple domain walls also introduces some crossover time dependence in $d_{\rm cont}$, accounting for its eventual linear scaling in $L$ when $T \gg \exp\lz \beta \sigma L \rz$.
Indeed, the decoupling conditions Eqs.~(\ref{eq:decoupling_condition}, \ref{eq:decoupling_condition_estimate}) retain their forms in this limit, 
and $d_{\rm cont}$ can still be identified with $|A|^\ast$, which approaches $L/2$ in the long time limit.

%Notice that at the critical point, the crossover is accounted for by a conformal mapping.

\section{Numerical results~\label{sec:numerics}}

In this section, we compare our capillary-wave theory results with numerical computations in the random Clifford circuit (Fig.~\ref{fig:upcircuit}) for the observables explored in the previous section.
Overall, we find qualitative agreement between the two,
but as we shall see, a complete quantitative agreement is lacking.
%It is possible that a domain wall picture still holds, but the fluctuations are not identitical to thermal fluctuations of Ising domain walls.
We interpret the former as support for the general entanglement domain wall picture, and the latter as an indication of a more complex nature of these domain walls for the random Clifford circuit. %, possibly described by a generalized version of capillary-wave theory.

%\subsection{Decoupling of domain walls and scaling of the contiguous code distance \label{sec:clf_numerics_decoupling} }

%----------------------------
\begin{figure}[t]
    \centering
    \includegraphics[width=.48\textwidth]{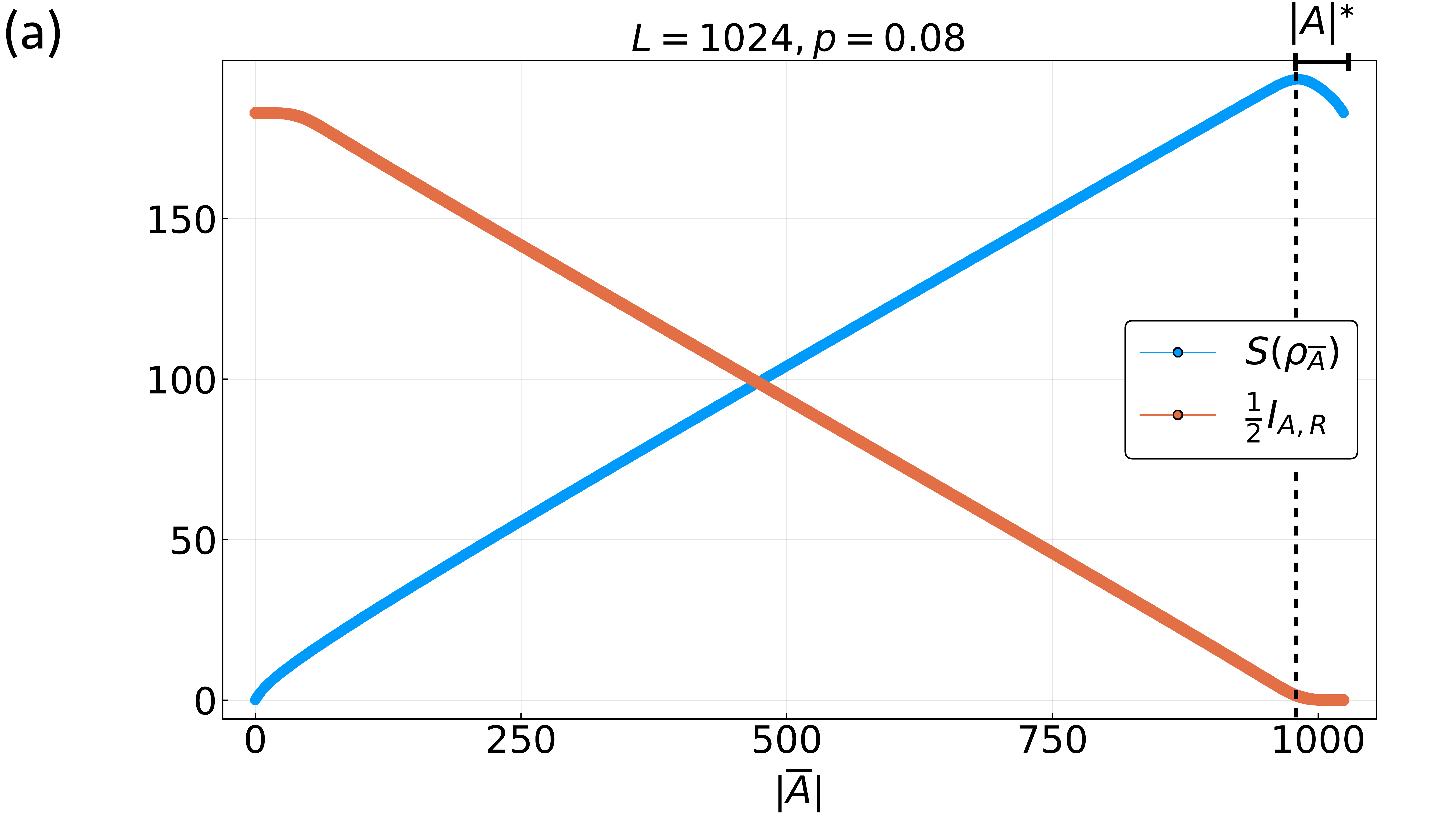}
    \includegraphics[width=.48\textwidth]{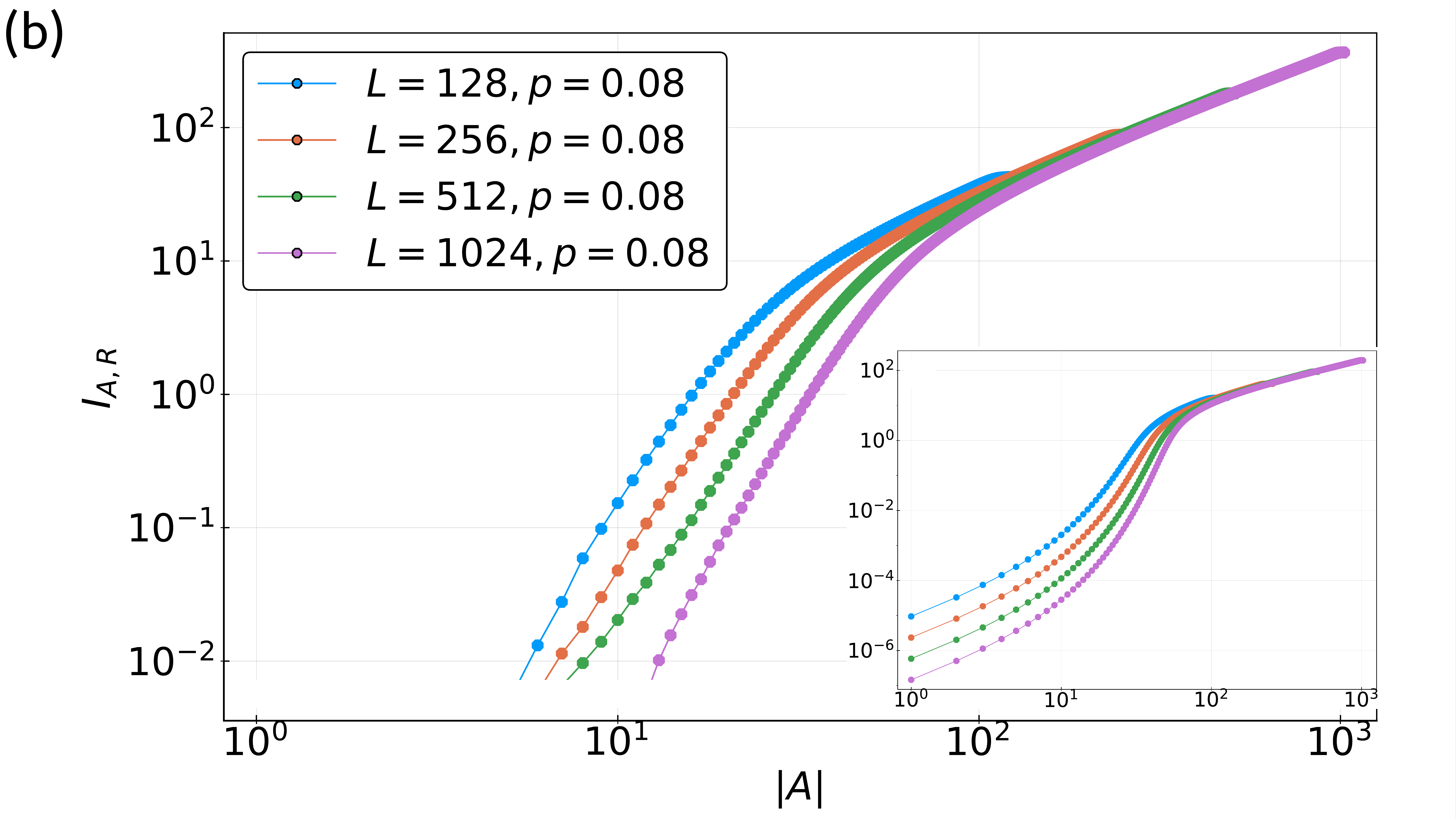}
    \includegraphics[width=.48\textwidth]{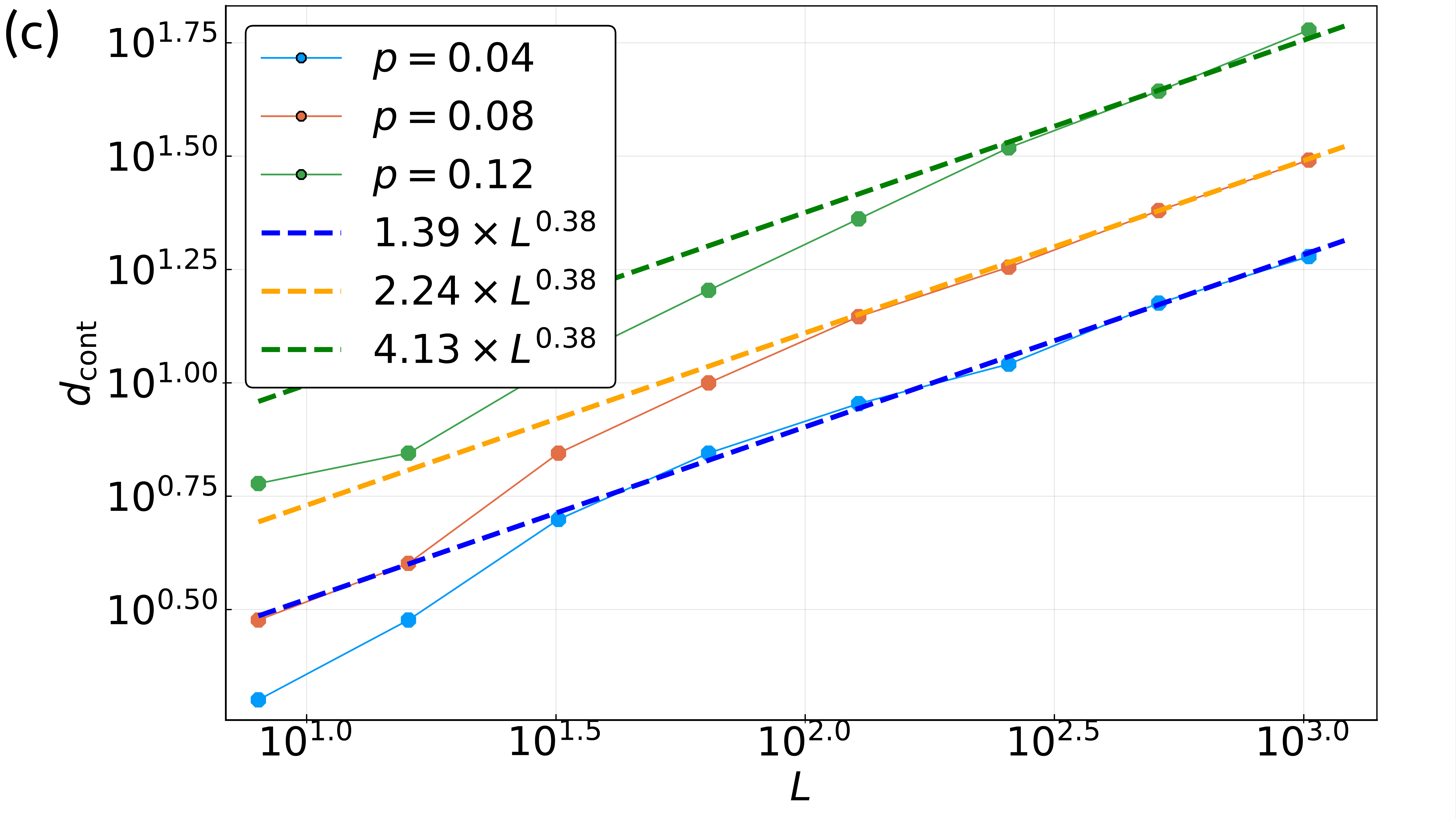}
\caption{
(a) $S(\rho_{\ovl{A}})$ and $\frac{1}{2} I_{A,R}$ from random Clifford circuit numerics, where we observe qualitative agreements with capillary-wave theory (Fig.~\ref{fig:bc_decouple}(c)).
(b) A closer look at $I_{A,R}$ on a log-log scale, where we find qualitative agreement, within accessible numerical resolutions, with the capillary-wave theory result (inset), computed from Eqs.~(\ref{eq:s_rho_A_small}, \ref{eq:s_rho_Q}, \ref{eq:decoupling_condition_estimate}) at $\beta \sigma = 0.1$.
(c) The scaling of the code distance (obtained from (a,b) upon setting $\epsilon = \ln 2$) with the system size for $p = 0.04, 0.08, 0.12$, where we find $d_{\rm cont} \propto L^{\gamma_1}$ with $\gamma_1 \approx 0.38$.
}
\label{fig:clf_dec}
\end{figure}
%----------------------------

\subsection{Code distance for Clifford QECCs \label{sec:clifford_d_cont} }

The most striking qualitative prediction of the domain wall picture from Sec.~\ref{sec:dw_picture} is the phenomenon of decoupled domain walls as illustrated in Fig.~\ref{fig:bc_decouple}.
To explore this for random Clifford circuits, we compute $S(\rho_{\ovl{A}})$ and $I_{A, R}$ with varying $|A|$, taking a maximally-mixed initial state and averaging over the random ensemble of circuits, as well as over a time window $6L < T < 8L$.
Within this time window, we have $T \gg L^{1/2}$, so that Eqs.~(\ref{eq:s_rho_A_small}, \ref{eq:s_rho_Q}) should apply.
%With this particular generalization of capillary-wave theory, the scaling of $|A|^\ast$ in Eq.~\eqref{eq:Astar_vs_L} is modified to 
%\env{align}{
%    |A|^\ast \propto L^\gamma,
%}
%whereas the form of Eq.~\eqref{eq:decoupling_condition_estimate} remains unchanged.

Our numerical results are shown in Fig.~\ref{fig:clf_dec}(a) for $L = 1024$ and $p = 0.08 \approx 0.5 p_c$.  Strikingly, we observe the same non-monotonicity in $S(\rho_{\ovl{A}})$, decreasing with
$\ovl{A}$ in the range $L - |A|^\ast < |\ovl{A}| \le L$, 
in accordance with the capillary-wave theory results in Fig.~\ref{fig:bc_decouple}.
Moreover, within this range, $I_{A,R}$ is very small, showing a plateau with height $\approx 0$.
The Clifford numerical results for both $S(\rho_{\ovl{A}})$ and  $I_{A,R}$
are thus fully consistent with the domain wall decoupling results in Fig.~\ref{fig:bc_decouple}. Evidently, 
the domain wall picture holds for random Clifford circuits, being qualitatively consistent with capillary-wave theory.

Our particular choice of $p=0.08$ was unimportant.
Indeed, for the Clifford circuit we find consistency with Fig.~\ref{fig:bc_decouple} for a wide range of $p$ with $0 < p < p_c$ (not shown).
This is as expected, since the domain wall picture should be valid at any ``temperature'' $p$ below the ``critical temperature'' $p_c$ of the spin model, i.e. throughout the ``ordered phase''.

We also explore finite size effects on $I_{A,R}$ (see Fig.~\ref{fig:clf_dec}(b)).
For a fixed $|A| < |A|^\ast$, we find that $I_{A,R}$ decreases with increasing system size $L$, consistent with Eq.~\eqref{eq:decoupling_condition_estimate}.
We therefore expect that in the thermodynamic limit, $I_{A,R} = 0$ if and only if $|A| < |A|^\ast$; the identification between $|A|^\ast$ and $d_{\rm cont}$ can then be made.
In a finite system, we define $|A|^\ast$ as the size of $A$ for which $I_{A, R} \approx \epsilon$, where $\epsilon $ is a small number independent of $L$.

With this identification, we may now examine how $d_{\rm cont} = |A|^\ast$  depends on the system size (as obtained from Fig.~\ref{fig:clf_dec}(a,b)).
As shown in Fig.~\ref{fig:clf_dec}(c), we find that the code distance $d_{\rm cont}$ increases with increasing $p$, qualitatively consistent with capillary-wave theory.
The code distance also grows with $L$, but as a power-law function, $d_{\rm cont} \propto L^{\gamma_1}$.
The exponent is estimated to be $\gamma_1 \approx 0.38$, in agreement with a direct computation (from the algebraic definition of $d_{\rm cont}$) in Ref.~\cite{gullans1905purification}.
The power-law scaling of $d_{\rm cont}$ quantitatively differs from capillary-wave theory in Eq.~\eqref{eq:Astar_vs_L}, where a logarithmic scaling was found.

%----------------------------
\begin{figure}[t]
    \centering
    \includegraphics[width=.48\textwidth]{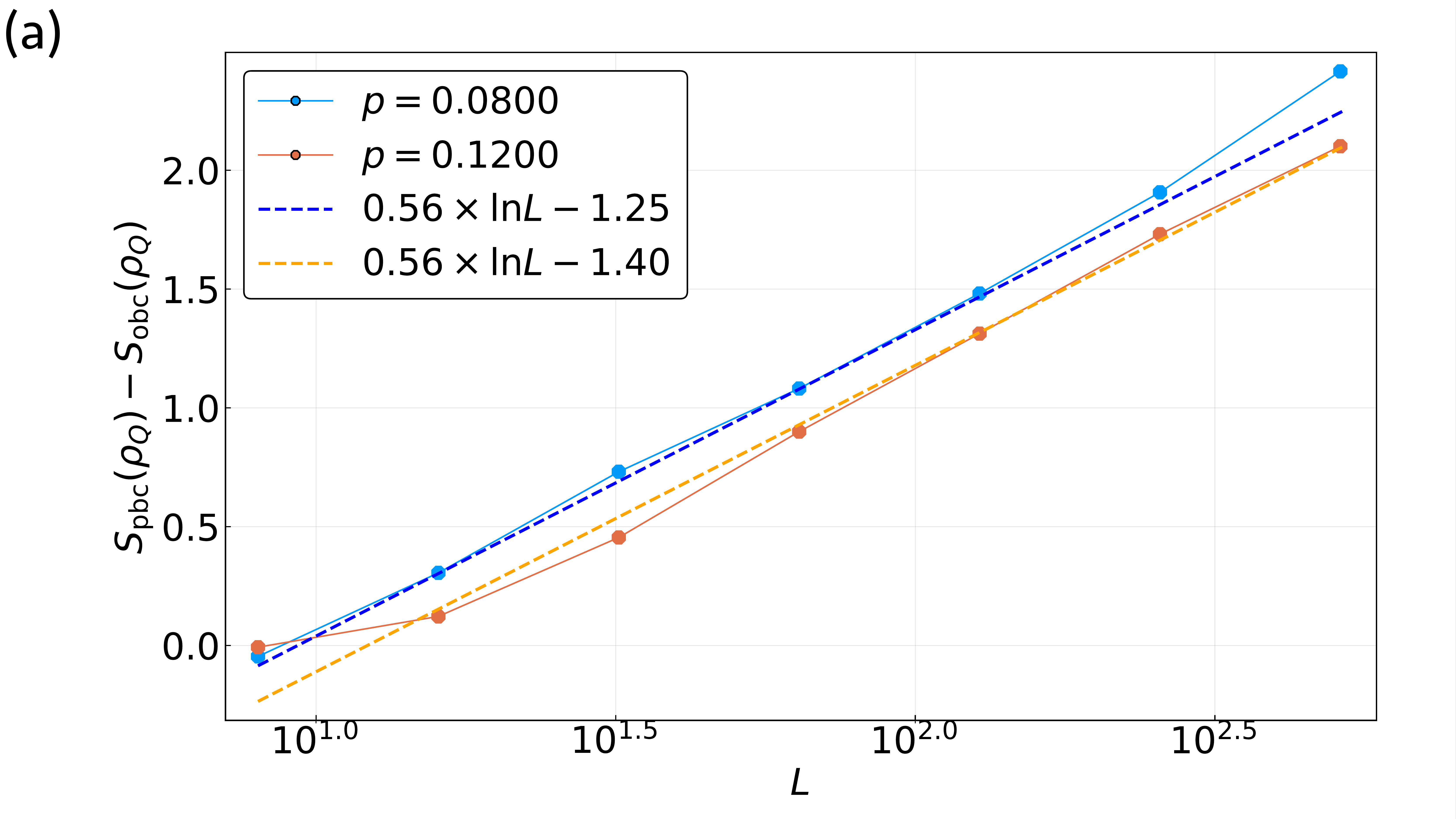}
    \includegraphics[width=.48\textwidth]{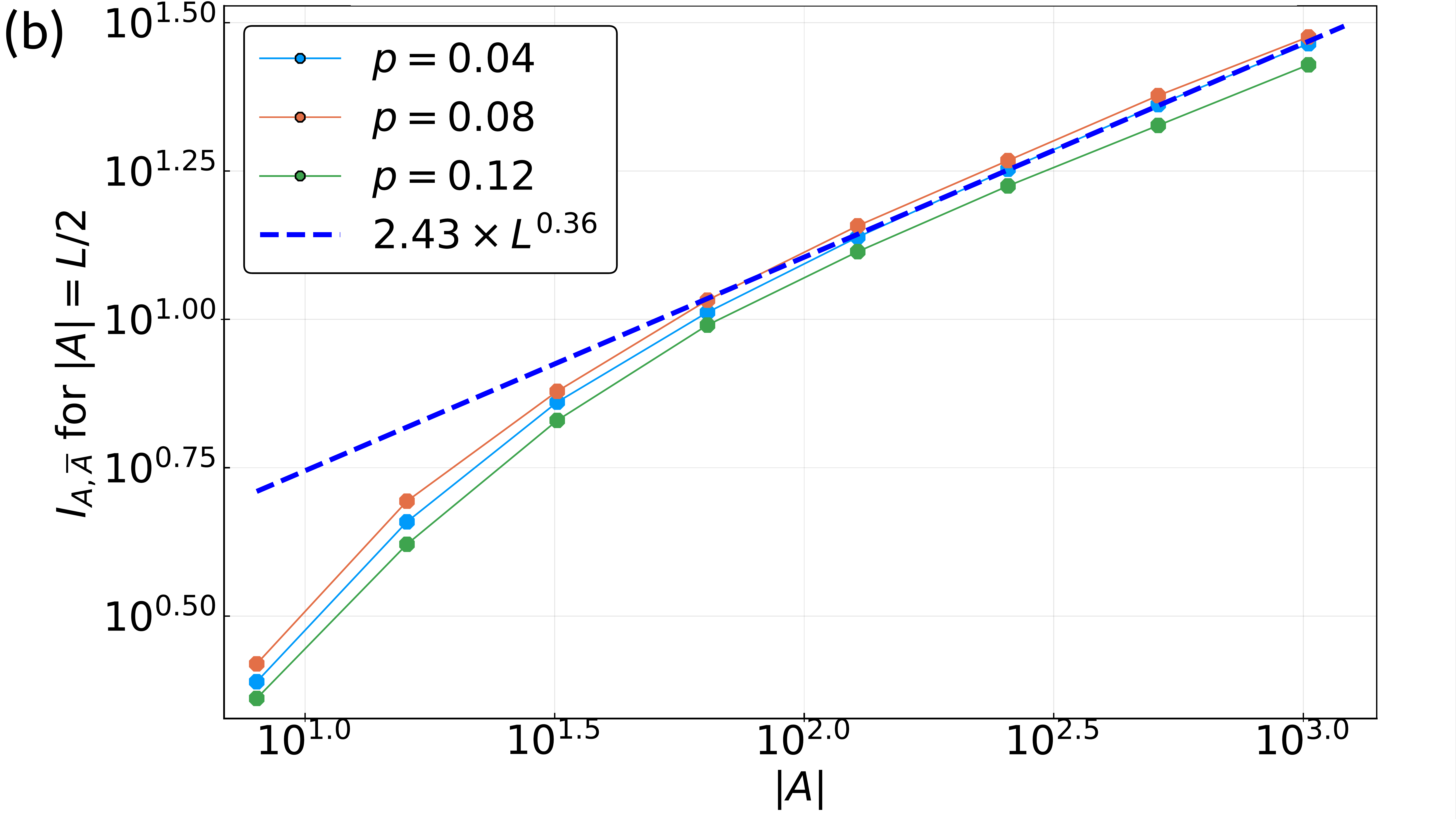}
\caption{
(a)
The difference between entropies of the entire system as computed for the random Clifford circuit with periodic and open boundary conditions.
We observe a logarithmic dependence on the system size,
$S_{\rm pbc}(\rho_Q) - S_{\rm obc}(\rho_Q) = \zeta \ln L$, with $\zeta \approx 0.56$.
This difference has a weak time depedence, but not displayed here.
(b)
The halfcut mutual information $I_{A, \ovl{A}}$ for $|A| = |\ovl{A}| = L/2$ as a function of $L$, with periodic b.c., where we find $I_{A, \ovl{A}} \propto L^{\gamma_2}$ with $\gamma_2 \approx 0.36$.
}
\label{fig:endpoint_entropy}
\end{figure}
%----------------------------

\subsection{Clifford dynamics versus (generalized) capillary-wave theory \label{sec:clifford_vs_GCW} }

We next numerically compute a few more quantities that we can compare with capillary-wave theory, as shown in Fig.~\ref{fig:endpoint_entropy}.
Once again, these results were obtained for the random Clifford circuit with a maximally-mixed initial state, upon averaging over both circuit realizations and the time window $6L < T < 8L$.

In Fig.~\ref{fig:endpoint_entropy}(a), we plot the difference between $S(\rho_Q)$ with periodic and open boundary conditions, $\Delta S(\rho_Q) \coloneqq S_{\rm pbc}(\rho_Q) - S_{\rm obc}(\rho_Q)$.
Capillary-wave theory (Eqs.~(\ref{eq:s_rho_A_small}, \ref{eq:s_rho_Q})) predicts cancellations of the time dependence as well as of the ``surface energy'' term, leaving only the extra endpoint entropy term, $(1/2) \ln L$.
This logarithimic scaling is indeed observed numerically, with the coefficient of $\ln L$  given by $\approx 0.56$, close in value to that of capillary-wave theory.
We have also confirmed a very weak $T$-dependence of $\Delta S(\rho_Q)$ on intermediate time scales, but the data is not displayed here.

In Fig.~\ref{fig:endpoint_entropy}(b), we plot the ``halfcut mutual information''~\cite{gullans1905purification}, $I_{A, \ovl{A}}$ with $|A| = |\ovl{A}| = L/2$ versus $L$ with periodic b.c..  Upon varying $L$, we find $I_{A, \ovl{A}} \propto L^{\gamma_2}$ with $\gamma_2 \approx 0.36$, and the overall amplitude having a weak dependence on $p$.
%This behavior is captured by the postulated GCW theory (compare Eqs.~(\ref{eq:F_GCW_A}, \ref{eq:F_GCW_Q})), and the estimate $\gamma \approx 0.36$ is close in value to the estimate as obtained from $d_{\rm cont}$ (Fig.~\ref{fig:clf_dec}(c)).
As for the code distance in Fig.~\ref{fig:clf_dec}(c) which grows with a similar power $\gamma_1 \approx \gamma_2$, this power-law scaling is quantitatively different from capillary-wave theory. The latter predicts a logarithmic scaling, $I_{A, \ovl{A}} = (7/2) \ln L$ for $T \propto L$.

To account for the power-laws in Fig.~\ref{fig:clf_dec}(c) and Fig.~\ref{fig:endpoint_entropy}(b), we introduce a phenomenological description, which we call ``generalized capillary-wave'' (GCW),
with the following (minimal) modifications of the free energies for ``pinned'' and ``waist'' domain walls (Eqs.~(\ref{eq:s_rho_A_small}, \ref{eq:s_rho_Q})), respectively
\env{align}{
    \label{eq:F_GCW_A}
    &\  F_{\rm GCW} (A) \nn
    =&\ \beta \sigma |A| + \chi |A|^\gamma, \text{ when } T \gg L^{\zeta} \gg |A|^\zeta, \\
    \label{eq:F_GCW_Q}
    &\  F_{\rm GCW} (Q) \nn
    =& \env{cases}{
    \beta \sigma L -\ln T% + \chi_Q L^\beta
    , & \text{open b.c.} \\
    \beta \sigma L -\ln \frac{T}{L^\zeta}%+ \chi_Q L^\beta
    , & \text{periodic b.c.}
    }
    \text{ when } T \gg L^{\zeta}.
}
Here $0 \le \gamma < 1$ is the exponent characterizing domain wall free energies in GCW, and $0 < \zeta < 1$ is the exponent of vertical extent of the domain walls.\footnote{Notice that with $0 < \zeta < 1$, we still have $T \gg L^\zeta$ for $6L < T < 8L$, the time window we took in the numerics.
As we saw in Fig.~\ref{fig:endpoint_entropy}(a), the exponent $\zeta \approx 0.56$ falls within this range, and seems to be close in value to $\zeta_{\rm CW} = 1/2$.
}
The constant $\chi$ is expected to be independent of $|A|, L, T, \beta \sigma$.
Capillary-wave theory thus has $\gamma_{\rm CW} = 0$ and $\zeta_{\rm CW} = \frac{1}{2}$; compare Eqs.~(\ref{eq:s_rho_A_small}, \ref{eq:s_rho_Q}).
This generalization of capillary-wave theory remains qualitatively consistent with Fig.~\ref{fig:clf_dec}(a,b) and Fig.~\ref{fig:endpoint_entropy}(a), where we found $\zeta \approx 0.56$.

Eqs.~(\ref{eq:F_GCW_A}, \ref{eq:F_GCW_Q}), together with the %reasoning
definition of $|A|^\ast$ in Eq.~\eqref{eq:Astar_vs_L},\footnote{We note that the form of Eq.~\eqref{eq:decoupling_condition_estimate} is identical for capillary-wave theory and its generalization in Eqs.~(\ref{eq:F_GCW_A}, \ref{eq:F_GCW_Q}), since in its derivation we only kept the leading linear term, which is common for both cases.
Thus the identification between $|A|^\ast$ and $d_{\rm cont}$ can still be made for GCW.} lead to the following scaling behaviors for the code distance and half-cut mutual information,
%modification of %Eq.~\eqref{eq:Astar_vs_L}
\env{align}{
    d_{\rm cont} =&\ |A|^\ast \approx \frac{\chi}{2\beta \sigma} L^\gamma, \\
    I_{A, \ovl{A}} \propto&\ L^\gamma . 
}
These are both consistent with our Clifford numerics in Fig.~\ref{fig:clf_dec}(c) and Fig.~\ref{fig:endpoint_entropy}(b), provided we take $\gamma = \gamma_1 = \gamma_2$.

We emphasize that Eqs.~(\ref{eq:F_GCW_A}, \ref{eq:F_GCW_Q}) are phenomenological, motivated by both capillary-wave theory and our Clifford numerics (specifically Fig.~\ref{fig:clf_dec}(c) and Fig.~\ref{fig:endpoint_entropy}(b)). %, which are convenient to have in order to organize our presentation;
At this moment we do not have a theory from which these free energies can be derived.
%Moreover, their forms are not unique, but are chosen to fit the numerics, and happen to be simple.

%For this reason, we will \emph{not} present fitting to  capillary-wave theory.
%We will be more detailed when these disagreements are encountered; these issues are further discussed in Sec.~\ref{sec:discussion}.
%Indeed, there is no reason that ``entanglement domain walls'' in the random Clifford circuit (if they exist) are exactly of the simplest type; on the contrary, it is very likely that entanglement domain walls in random Clifford circuit, and in the more general class of circuits, 
%Previously in Refs.~\cite{li1901hybrid, gullans1905purification}, qualitative agreements between Eqs.~(\ref{eq:s_rho_A_small}, \ref{eq:s_rho_Q}) and random Clifford circuit numerics were found.
%However, we will not discuss these observables, 

%In Sec.~\ref{sec:clf_numerics_decoupling} and Sec.~\ref{sec:clf_numerics_d_cont_scaling}, we focus on the decoupling criterion and the scaling of the code distance. %, which are the most relevant criteria of the dynamically generated state as a QECC.
%In Sec.~\ref{sec:clf_numerics_endpoint_entropy}, we compare endpoint entropy for ``waist domain walls'' in open and periodic boundary conditions as in Fig.~\ref{fig:bc_waist}.

%\subsection{Comment on a direct comparison of $S(\rho_A)$ (for $|A|\ll L$) and $S(\rho_Q)$ with capillary-wave theory}

We note that direct numerical computations of $S(\rho_A)$ (for $|A|\ll L$) and $S(\rho_Q)$ are qualitatively consistent with both capillary wave theory (Eqs.~(\ref{eq:s_rho_A_small}, \ref{eq:s_rho_Q})) and its generalization in  Eqs.~(\ref{eq:F_GCW_A}, \ref{eq:F_GCW_Q}), as established in Refs.~\cite{li1901hybrid, gullans1905purification}.
In particular, 
\env{itemize}{
\item
For the approximation $S(\rho_A) \approx F_{\rm CW}(A)$ when $|A| \ll L$,
the ``linear plus log'' form of $F_{\rm CW}(A)$ is consistent with the stabilizer length distribution~\cite{li1901hybrid};
\item
For the approximation $S(\rho_Q) \approx F_{\rm CW}(Q)$,
the $-\ln T$ dependence on circuit depth is consistent with an exponentially long purification time~\cite{gullans1905purification}.
}
On the other hand, a quantitative comparison between capillary-wave theory and GCW is tricky, due to the difficulty in distinguishing a logarithimic function from a small power-law in the presence of a background linear term.
%both quantities involve a ``linear plus sublinear'' form, and there is no simple way to subtract off the linear ``surface energy'' term.
%One can of course tune the surface tension $\beta \sigma$ for a best fit; but this best fit will not work for other observables.
Thus, we will not here attempt to compare capillary-wave theory and GCW for the quantities $S(\rho_A)$ (with $|A|\ll L$) and $S(\rho_Q)$.

%Since qualitative agreements for $S(\rho_A)$ (for $|A|\ll L$) and $S(\rho_Q)$ are well established in Refs.~\cite{li1901hybrid, gullans1905purification}, we do not present the (rather boring) plots here.

%\subsection{Endpoint entropy}

%In Fig.~\ref{fig:endpoint_entropy}, we plot the difference between $S(\rho_Q)$ in periodic and open boundary conditions, $\Delta S(\rho_Q) \coloneqq S_{\rm pbc}(\rho_Q) - S_{\rm obc}(\rho_Q)$.
%Either capillary-wave theory (Eqs.~(\ref{eq:s_rho_A_small}, \ref{eq:s_rho_Q})) or its generalization (Eqs.~(\ref{eq:F_GCW_A}, \ref{eq:F_GCW_Q})) would predict cancellations of the time dependence as well as of the ``surface energy'' term, leaving only the extra endpoint entropy term $\zeta \ln L$.
%This is indeed observed numerically, where we find $\zeta \approx 0.56$.
%We have also confirmed a very weak $T$-dependece of $\Delta S(\rho_Q)$, but the data is not displayed here.
%The value of $\zeta$ is close in value to that of the simple capillary-wave theory $\zeta_{\rm CW} = 1/2$, perhaps fortuitously.

\section{Discussion~\label{sec:discussion}}

\subsection{Summary}

In this paper  
we established a correspondence between QECCs generated by random hybrid Clifford circuit dynamics, and the statistical mechanics of fluctuating ``entanglement domain walls''.
The number of encoded logical qubits $k$ of the QECC maps to the ``surface energy'' that is extensive in the number of physical qubits $|Q|$, and the code distance maps to a crossover length scale proportional to the ``entropy'' of transverse fluctuations, that is subextensive in $|Q|$.
Fluctuations of entanglement domain walls are entirely responsible for the diverging code distance, which protects the state against local (undetectable/uncorrectable) errors, a characteristic property of QECCs.

Our results rest upon two well-motivated assumptions, namely the validity of the entanglement domain wall picture, and the ``linear plus sublinear'' form of their free energies.
The former has been analytically established in the context of hybrid random Haar circuits~\cite{andreas2019hybrid, choi2019spin},
and the latter follows from the former within capillary-wave theory. We expect that both assumptions are also valid for Clifford circuits, as supported by the Clifford numerics in Refs.~\cite{li1901hybrid, gullans1905purification}, as well as those in Sec.~\ref{sec:numerics}.

We emphasize that 
the qualitative properties of the QECC do not depend crucially on the specific form of the entropic term, which diverges logarithmically with $|Q|$ in capillary-wave theory, and as a small power-law in our Clifford numerics. %  the only requirement being that it should diverge with $|Q|$, and scales subextensively with $|Q|$.
%This is exemplified by capillary-wave theory (where the entanglement domain walls are treated as Ising like) which we use as an illustrative tool, and is confirmed by numerical results (in Sec.~\ref{sec:numerics}), that are 
The latter is possibly described by a certain generalization of capillary-wave theory.
In some sense, one can view capillary-wave theory as a ``mean-field theory'' of the entanglement domain walls.

\subsection{The diverging code distance as a self-consistency condition}

The error correcting nature of the dynamically generated state, as exemplified by the diverging code distance, is consistent with the resilience of this finite entropy-density (and code rate) state to repeated local measurements.
Indeed, a measurement decreases the entropy only if the measured operator is a nontrivial logical operator (see Sec.~\ref{sec:qecc_dynamics}), and with a diverging code distance
the probability of each local measurement in the circuit (Fig.~\ref{fig:upcircuit}) being a logical operator (denoted $p_{\rm logical}$) vanishes in the thermodynamic limit (equivalently, each qubit in $Q$ decouples from the reference state $R$ with probability one).
We can estimate $p_{\rm logical}$ by setting $|A| = 1$
in Eq.~\eqref{eq:decoupling_condition_estimate}, giving $p_{\rm logical} \propto \ell_A = I_{A,R} \approx \exp \lz -2 \beta \sigma d_{\rm cont} \rz $, leading to $p_{\rm logical} \propto L^{-2}$ within capillary-wave theory for $T \propto L$, and $p_{\rm logical} \propto \exp\lz -\chi L^\gamma \rz$ within a generalized capillary-wave description.  In either case, this
leads to a vanishing rate of purification in the mixed phase when $O(L)$ measurements are made in each time step~\cite{gullans1905purification}, and subsequently to the stability of the finite code rate.\footnote{When $T$ is allowed to be independent of and longer than $L$, capillary-wave theory gives,
\env{align}{
    p_{\rm logical} \propto (LT)^{-1} \ \Rightarrow \ \frac{dS(\rho_Q)}{dT} \approx - (pL) p_{\rm logical} \propto T^{-1}, \nonumber %or, upon integrations $S_Q = \beta \sigma L - \ln(T)$,
}
consistent with Eq.~\eqref{eq:s_rho_Q} and an exponentially long purification time (see Sec.~\ref{sec:crossover}).
}

Our discussion above is {not} an explanation of the stability of the mixed phase, but a requirement of self-consistency, since the diverging code distance is itself computed from the steady state within the mixed phase.
%\footnote{In the area-law phase, the self-consistency condition is %satisfied in a different way.
%With $S(\rho_Q) = k = 0$, there are no logical operators, so the %measurement operator is never logical.}
The domain wall picture itself also requires the assumption of an ordered phase.

Moreover, a quantum Hamming bound~\cite{ekert1996hamming, gottesman9604hamming} on $p_c$, as in Ref.~\cite{fan2020selforganized}, cannot be inferred from our discussion.
Besides the code distance being subextensive rather than extensive,
here we are viewing the one-qubit measurements within one circuit time step as a sequence of one-qubit errors, rather than a single $p|Q|$-qubit error.
With respect to these single qubit errors, the code is highly degenerate, and the Hamming bound does not apply.
%\YL{Furthermore, we do not require the following unitary layer to correct these errors of logical operators; with our mixed state formulation, the errors induced by logical operators are not detectable in the first place.}

%----------------------------
\begin{figure}[t]
    \centering
    \includegraphics[width=.48\textwidth]{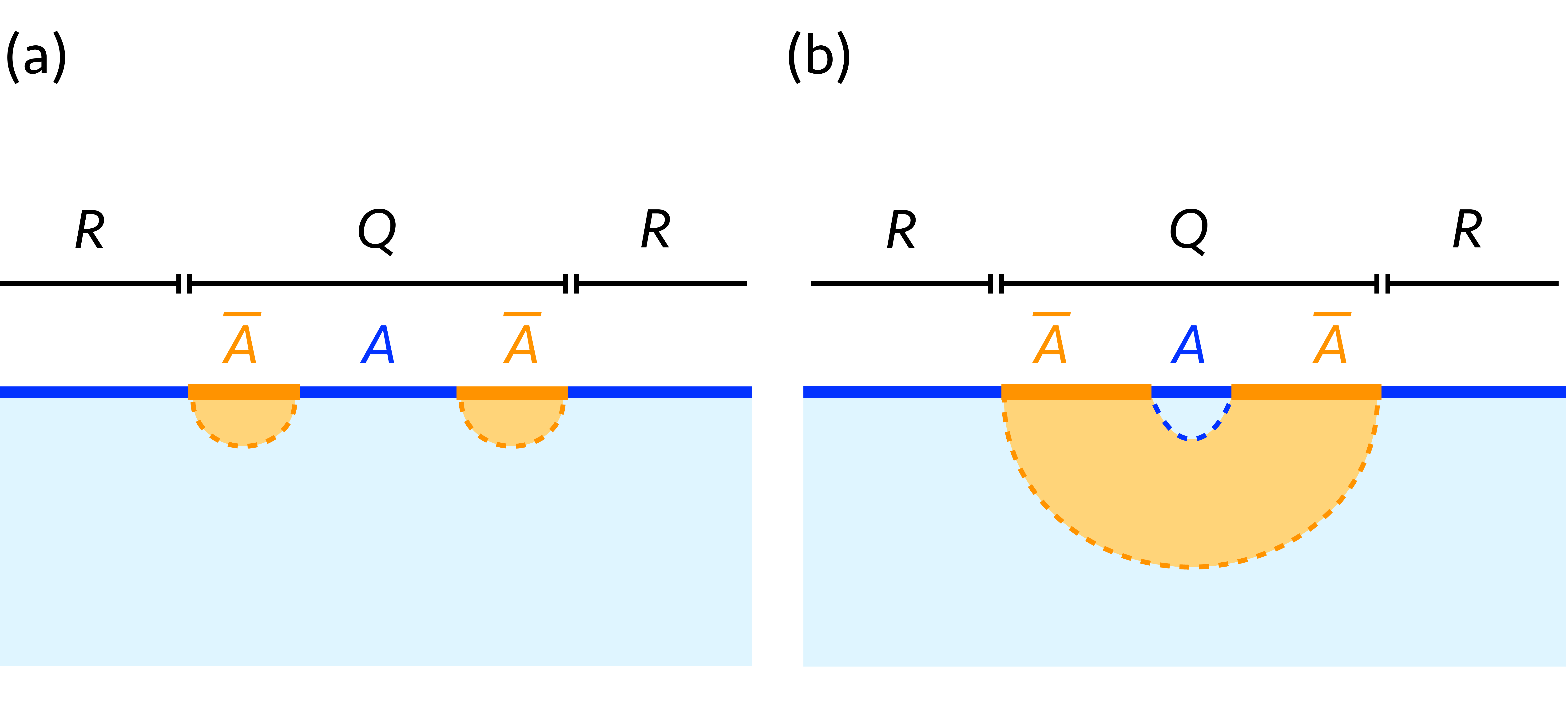}
\caption{
The circuit, when dynamically evolving an initial pure state, can also be incorporated in the QECC framework by taking an extensive subsystem $Q$ as the QECC, and the complement of $Q$ as the reference $R$, with $|Q| < |R|$.
When $A > d_{\rm cont}$, the dominant domain wall configuration is shown in (a), and in this regime, $A$ and $R$ have nonvanishing correlation.
On the other hand, when $A < d_{\rm cont}$ the dominant domain wall configuration is the ``rainbow diagram'' shown in (b), implying that $A$ and $R$ should fully decouple (for $|Q|$ large), with vanishing mutual information.
In this regime, an error on $A$ will have no effect on $S(\rho_Q)$.
}
\label{fig:bc_pure}
\end{figure}
%----------------------------

Finally, we mention that QECCs can also be dynamically generated for circuits with a pure initial state \YL{in the volume law entangled phase when $p < p_c$}~\cite{choi2019qec, fan2020selforganized}, if we take the ``system" $Q$ to be an extensive subsystem, $R$ to be the complement of $Q$ with $ |R|>|Q|$, and consider the decoupling of $A \subseteq Q$ from $R$ (see Fig.~\ref{fig:bc_pure}).
Indeed, in this case our Clifford numerics (not shown) demonstrate the presence of these decoupling conditions (e.g. a vanishing $I_{A,R}$ for $|A|<|A|^\ast = d_{\rm cont}$), qualitatively consistent with capillary-wave theory.
Other results within this setup should be similar to those obtained in Refs.~\cite{li1901hybrid, choi2019qec, fan2020selforganized}.
%\YL{We note a similarity between this discussion and those made in Ref.~\cite{li1901hybrid}, where the subleading }

\subsection{The role of disorder}

As for random Haar circuits~\cite{andreas2019hybrid, choi2019spin} and random tensor networks~\cite{vasseur2018rtn}, the identification in Eq.~\eqref{eq:SA_log_Z} is between free energies in the stat.~mech. model and entanglement entropies \emph{averaged} over an ensemble of circuits.
Thus we have been studying the averaged entropies, and comparing them with (generalized) capillary-wave theory.
%As discussed in Sec.~\ref{sec:clf_numerics_beyond_cw}, 
Capillary-wave theory assumes translational symmetry by construction, with no reference to sample-to-sample fluctuations or the role of disorder.

In Fig.~\ref{fig:clf_more}, we present the statistical sample-to-sample fluctuation of $S(\rho_A)$ over an ensemble of random Clifford circuits, versus the subregion size $|A|$, for $0 \le |A| \le L/2$.
Previously in Ref.~\cite{li1901hybrid}, the distribution of $S(\rho_A)$ was found to be Gaussian-like.
Here, we find the following power-law scaling for the standard deviation (square root of the variance) of the entropy,  $\sqrt{\mathrm{var}[S(\rho_A)]} \propto |A|^{0.33}$, with an amplitude that depends weakly on $p$.
This power-law behavior is interesting, yet beyond \emph{any} generalization of capillary-wave theory, as the latter always describes a clean system, for which the notion of an ensemble of disorder realizations is irrelevant.
%However, the result in Fig.~\ref{fig:clf_more} shows that the variance of $S(\rho_A)$ over disorder realizations is divergent with $|A|$ as a power-law, $|A|^{0.33}$, so
This result suggests that %the effects of disorder diverge with $|A|$, and 
disorder could dramatically modify the structure of the domain walls, possibly accounting for the power-law dependences in $d_{\rm cont}$ and $I_{A, \ovl{A}}$ in Sec.~\ref{sec:numerics}. %, as well as the critical properties at $p_c$.
%It is therefore no surprise that the entanglement domain walls are not \emph{exactly} described by capillary-wave theory.
%and the generalized capillary-wave theory of average free energies proposed in Sec.~\ref{sec:numerics} could likely be a consequence of disorder.

%----------------------------
\begin{figure}[t]
    \centering
    \includegraphics[width=.48\textwidth]{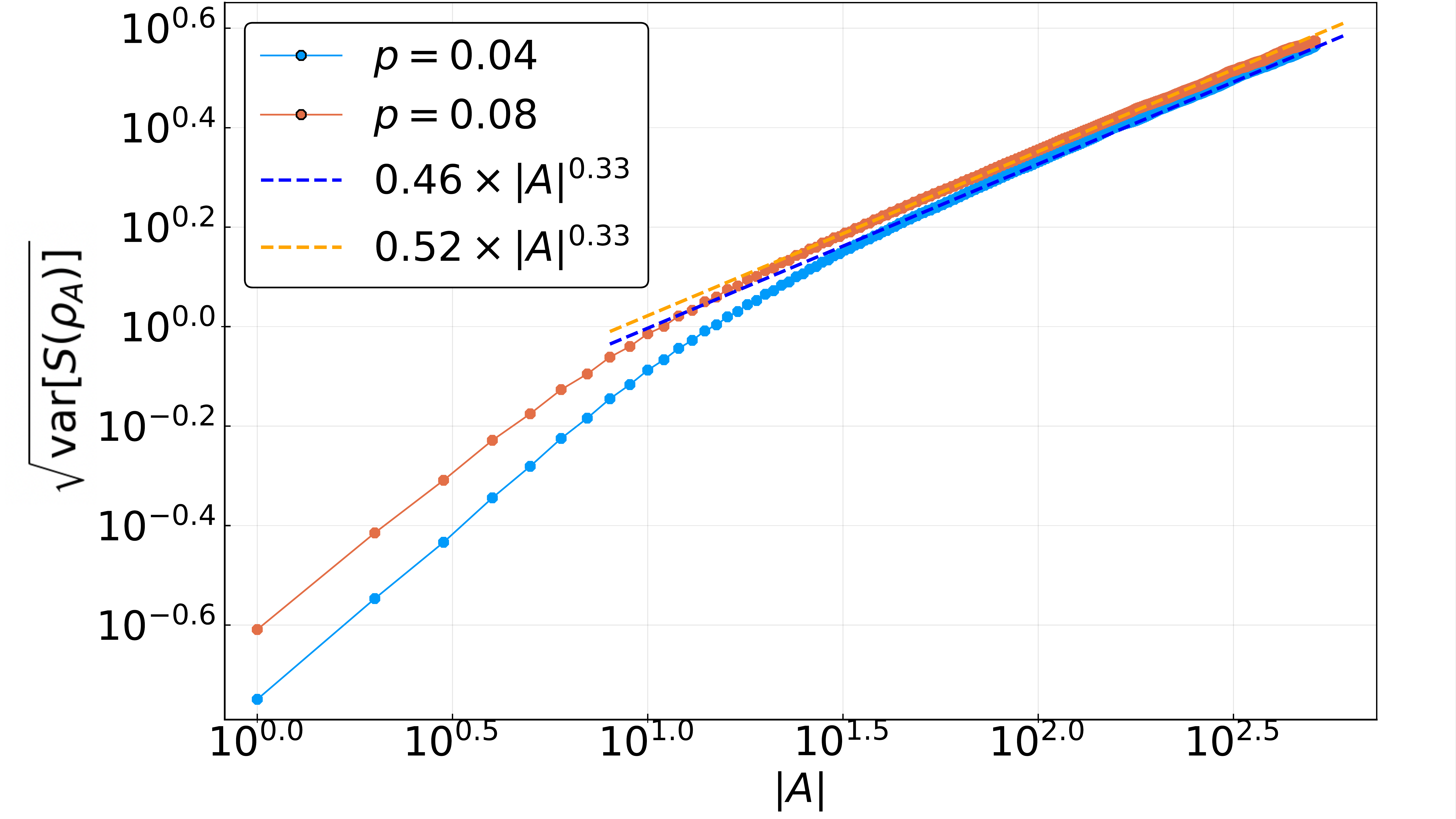}
\caption{The sample-to-sample fluctuation of $S(\rho_A)$ as a function of $|A|$, obtained from an ensemble of random Clifford circuits.
We take $L = 1024$ and $0 \le |A| \le L/2$.
}
\label{fig:clf_more}
\end{figure}
%----------------------------

We remark that the exponent for the standard deviation $0.33$, as well as the exponent $\gamma \approx 0.36$, are both close to the exponent $1/3$ for subextensive corrections to free energies of a directed polymer in random media (DPRM)~\cite{huse_henley_1985_DPRM, kardar1985roughening, huse_henley_fisher_1985_respond, kardar1987DPRM} that falls within the Kardar--Parisi--Zhang (KPZ) universality class~\cite{KPZ1986}. Such corrections are due to quenched disorder.
We note that similar scaling behaviors have been found in random unitary circuits \emph{without} measurements~\cite{nahum2017KPZ, nahum2018operator, zhou1804emergent, zhou1912membrane}.
Without an analytic theory, we cannot determine if the entanglement domain walls are indeed DPRM-like.  In this context, 
it could be interesting to find ``clean'' circuit models, for which effects of quenched disorder are absent, so that the subleading ``entropic'' term only receives contribution from thermal fluctuations, just like simple Ising domain walls.
These open issues are left for future work.

\subsection{Outlook}

We have made extensive use of the random Clifford circuit and the stabilizer formalism to establish our results.
%, we note that there are non-stabilizer QECCs~\cite{RHSS1997nonadditive}, and we expect that the entanglement domain wall picture should be valid in those cases as well
It would be interesting to see if the entanglement domain wall picture is valid in a broader class of models, %beyond random Clifford circuits,
such as hybrid random Haar circuits~\cite{nahum2018hybrid, choi2019spin, andreas2019hybrid, huse1911tripartite, harrow2001efficient} or other ``generic'' models of non-unitary dynamics~\cite{cao2018monitoring, szyniszewski1903measurement, Tang2019, chamon2001nonuniversal, danshita2001cold, chenxiao2004nonunitary, ashida2004continuous, diehl2005trajectory, pal_lunt_2005_mbl_hybrid, buechler2006projectiveTFIM}, or circuit models with measurements only~\cite{nahum1911majorana, barkeshli2004symmetric, hsieh_sang_2004_protected, ippoliti2004measurementonly}.

While a general QECC does not necessarily have ``locality'', spatial locality and spatial dimensionality is important in the hybrid circuit dynamics.
For this reason, we have been considering the ``contiguous code distance'', rather than the conventionally defined code distance in the QECC context.
We have also restricted our attention to (1+1)-dimensional circuits, in which the domain walls are one-dimensional objects in a two-dimensional background.
It would be interesting to test this picture in higher dimensions~\cite{nahum2018hybrid}, %, dalmonte2007twoplusonedim}, %~\cite{harrow2001efficient},
in tree-like geometries~\cite{vasseur2003mft}, or even in ``all-to-all'' models~\cite{gullans1905purification, sagar2005volume} where locality is entirely absent.

The dynamically generated QECCs are found to have a finite code rate and a subextensive code distance.
They are therefore not ``good codes'' in the conventional sense~\cite{calderbank9512good}, which have a finite code rate and an extensive code distance.
For example, a good code can be obtained by running a random unitary circuit without measurements into the steady state, starting from a mixed state with a finite entropy density~\cite{haydenpreskill0708}.
It would be interesting to see
%It is not clear if the dynamically generated codes are of practical use.
%It is also not clear whether they are representative of a class of quantum states.
%These are questions worth exploring in the future.
%Going beyond hybrid circuits, one can ask
if the domain wall picture sheds any light on these good codes, and more generally on QECCs that are not dynamically generated and/or stabilizer-based~\cite{RHSS1997nonadditive}, e.g. those commonly used in an error correction context (see for example Ref.~\cite{brun1910survey} for a recent survey).

\iffalse
The dynamically generated QECCs are found to have a finite code rate and a subextensive code distance.
They are therefore not ``good codes'' in the conventional sense (which have a finite code rate and an extensive code distance)~\cite{calderbank9512good}.
However, we note that a domain wall picture -- though of a different kind -- also applies to a class of ``random good codes'', as generated by a deep ($T \gg L$) random unitary circuit \emph{without} measurements, starting from a mixed initial state at a finite entropy density smaller than $\ln 2$.
Entanglement entropies in such a circuit can be approximated by ``minimal cuts'' in the underlying lattice with no broken bonds, following constructions in Refs.~\cite{nahum2017KPZ, nahum2018hybrid, li2003cft}.
It would be interesting to see
%It is not clear if the dynamically generated codes are of practical use.
%It is also not clear whether they are representative of a class of quantum states.
%These are questions worth exploring in the future.
%Going beyond hybrid circuits, one can ask
if the domain wall picture sheds light on a more general class of good codes, e.g. those not dynamically generated and/or stabilizer-based~\cite{RHSS1997nonadditive}, as commonly used in an error correction context (see for example Ref.~\cite{brun1910survey} for a recent survey).
\fi

%------------------------------------------------------------

\section*{Acknowledgements}

%\YL{Do you want to acknowledge Sagar and/or Everett?}
%YES!!!!

We thank Xiao Chen, Michael Gullans, David Huse, {Adam Nahum}, Sagar Vijay, Yi-Zhuang You, Tianci Zhou, and especially Andreas Ludwig for helpful discussions.
We also thank Xiao Chen, Andrew Lucas, and Andreas Ludwig for collaborations on related topics.
{This work was supported by the Heising-Simons Foundation (YL and MPAF),
and by the Simons Collaboration on Ultra-Quantum Matter, which is a grant from the Simons Foundation (651440, MPAF).}
Use was made of computational facilities purchased with funds from the National Science Foundation (CNS-1725797) and administered by the Center for Scientific Computing (CSC). The CSC is supported by the California NanoSystems Institute and the Materials Research Science and Engineering Center (MRSEC; NSF DMR-1720256) at UC Santa Barbara.

%------------------------------------------------------------

\appendix

\begin{widetext}

\section{More on stabilizer codes \label{app:proof}}

\subsection{Entanglement entropies of stabilizer codes}

Let the stabilizer group $\mc{S}$ be an abelian subgroup of $\mc{P}(Q)$ as defined in Eq.~\eqref{eq:s_def}.
Let $\rho_Q(\mc{S})$ be the corresponding stabilizer code state as in Eq.~\eqref{eq:rho_S}~\cite{Fattal2004stabilizer},
\env{align}{
    \rho_Q(\mc{S}) = 2^{-|Q|} \sum_{g \in \mc{S}} g.
}
We can directly compute its R\'{e}nyi entropies~\cite{Fattal2004stabilizer},
\env{align}{
    \label{eq:s_rho_Q_derivation}
    & (\ln 2)^{-1} S^{(n)}(\rho_Q(\mc{S})) \nn
    =& \frac{1}{1-n} \log_2 \mathrm{Tr} \lz \( \rho_Q(\mc{S}) \)^n \rz \nn
    %=& - \log_2 \mathrm{Tr} \lz 2^{-2|Q|} \sum_{g, g^\p \in \mc{S}} g g^\p \rz \nn
    %=& - \log_2 \mathrm{Tr} \lz 2^{-2|Q|} |\mc{S}| \sum_{g \in \mc{S}} g \rz \nn
    =& \frac{1}{1-n} \log_2 \mathrm{Tr} \lz \(2^{-|Q|} |\mc{S}|\)^{n-1} \rho_Q(\mc{S}) \rz \nn
    =&\, |Q| - \log_2 |\mc{S}|.
}
%One can further verify that 
Since this result is independent of the R\'{e}nyi index $n$, we will suppress it henceforth.
As in Sec.~\ref{sec:setting}, we take
\env{align}{
    |\mc{S}| = 2^m
}
and define
\env{align}{
    k \coloneqq |Q| - m = (\ln 2)^{-1} S(\rho_Q(\mc{S})).
}

Given a bipartition of the system, $A \subseteq Q$, $\ovl{A} = Q - A$, we define the following group homomorphism
\env{align}{
    \label{eq:def_proj_Abar}
    \mathrm{proj}_{\ovl{A}}:\hspace{.6in} \mc{P}(Q) \quad \to &\quad \mc{P}(\ovl{A}) \nn
    g_A \otimes g_{\ovl{A}} \quad \mapsto&\quad g_{\ovl{A}}
}
We take the following reduced density matrix on $A$,
\env{align}{
    \label{eq:s_rho_A_reduced}
    &  \rho_A(\mc{S}) \nn
    =&\, \mathrm{Tr}_{\ovl{A}} \lz \rho_Q(\mc{S}) \rz\nn
    =&\, 2^{-|Q|} \sum_{g \in \mc{S}} \mathrm{Tr}_{\ovl{A}} (g) \nn
    =&\, 2^{-|A|} \sum_{g \in \mc{S} \cap \mathrm{Ker}\, \mathrm{proj}_{\ovl{A}}} g \nn
    =&\, 2^{-|A|} \sum_{g \in \mc{S}_A} g,
}
where we noticed that $\mathrm{Tr}_{\ovl{A}} (g)$ is nonzero only if $\mathrm{proj}_{\ovl{A}}(g) = \mathbbm{1}_{\ovl{A}}$, and defined $\mc{S}_A$ to be the following subgroup of $\mc{S}$,
\env{align}{
    \mc{S}_A \coloneqq \mc{S} \cap \mathrm{Ker}\, \mathrm{proj}_{\ovl{A}}.
}
Thus, we have
\env{align}{
    \mc{S}_A \cong &\ \frac{\mc{S}}{\mathrm{proj}_{\ovl{A}}(\mc{S})}, \nn
    |\mc{S}_A| =& \frac{| \mc{S}|}{| \mathrm{proj}_{\ovl{A} }(\mc{S}) |},
}
and from Eqs.~(\ref{eq:s_rho_Q_derivation}, \ref{eq:s_rho_A_reduced}),
\env{align}{
    \label{eq:s_rho_A_derivation}
    &  (\ln 2)^{-1} S(\rho_A(\mc{S})) \nn
    =& |A| - \log_2|\mc{S}_A| \nn
    =& |A| - \log_2 |\mc{S}| + \log_2 | \mathrm{proj}_{\ovl{A} }(\mc{S})|.
}

\subsection{Proof of Theorem 1 in Sec.~\ref{sec:theorem}}

Recall that $\mc{C}(\mc{S}) \subseteq \mc{P}(Q)$ is the abelianized centralizer of $\mc{S}$ in $\mc{P}(Q)$.
Recall also that the group of logical operators is defined as the quotient group $\mc{L} = \mc{C}(\mc{S}) / \mc{S}$ (see Sec.~\ref{sec:stabilizer}).
The homomorphism $\mathrm{proj}_{\ovl{A}}$ naturally induces the following homomorphism between quotient groups, 
\env{align}{
    \widetilde{\rm proj}_{\ovl{A}} :\hspace{.4in} \mc{L} \quad \to &\quad \frac{\mathrm{proj}_{\ovl{A}}(\mc{C}(\mc{S}))} {\mathrm{proj}_{\ovl{A}}(\mc{S})} \nn
    g \cdot \mc{S} \quad \mapsto&\quad 
    \mathrm{proj}_{\ovl{A}}(g) \cdot \mathrm{proj}_{\ovl{A}}(\mc{S})
}
It can be straightforwardly verified that this homomorphism is well defined, and %\footnote{That is, $[g] = [g^\p] \Rightarrow \widetilde{\rm proj}_{\ovl{A}}([g]) = \widetilde{\rm proj}_{\ovl{A}}([g^\p])$.}
%This induced homomorphism
is in fact surjective,
\env{align}{
    \widetilde{\rm proj}_{\ovl{A}}(\mc{L}) = \frac{ \mathrm{proj}_{\ovl{A}}(\mc{C}(\mc{S}))}{\mathrm{proj}_{\ovl{A}}(\mc{S})}.
}

Recall that the group $\mc{L}_A$ is defined in Sec.~\ref{sec:stabilizer} as follows
\env{align}{
    %\mc{L}_A \coloneqq \frac{\mc{C}(\mc{S}) \cap \(\mathrm{proj}_{\ovl{A}}\)^{-1} \lz \mathrm{proj}_{\ovl{A}}(\mc{S}) \rz} {\mc{S}},
    \mc{L}_A \coloneqq \frac{\ld g \in \mc{C}(\mc{S}) \ |\  \mathrm{proj}_{\ovl{A}}(g) \in \mathrm{proj}_{\ovl{A}}(\mc{S}) \rd} {\mc{S}},
}
where $\mathrm{proj}_{\ovl{A}}$ is understood as from $\mc{P}(Q)$ to $\mc{P}(\ovl{A})$, as in Eq.~\eqref{eq:def_proj_Abar}.
It follows from the definitions that
\env{align}{
    \mc{L}_A = {\rm Ker} \ \widetilde{\rm proj}_{\ovl{A}} \subseteq \mc{L}, 
}
thus
\env{align}{
    \label{eq:size_of_LA}
    & | \mc{L}_A | \nn
    =& 
    | {\rm Ker} \ \widetilde{\rm proj}_{\ovl{A}} | \nn
    =& \frac{| \mc{L} |}{| \widetilde{\rm proj}_{\ovl{A}}(\mc{L}) |} \nn
    =& \frac{| \mc{C}(\mc{S}) | \cdot | \mathrm{proj}_{\ovl{A}}(\mc{S}) |}{
    | \mc{S} | \cdot
    | \mathrm{proj}_{\ovl{A}}(\mc{C}(\mc{S})) |}.
}
In the following, we associate these factors with entanglement entropies, using Eq.~\eqref{eq:s_rho_A_derivation}.

We state without proof that an arbitrary generating set of $\mc{S}$ can be extended into one of $\mc{C}(\mc{S})$~\cite{calderbank1997quantum, aaronson0406chp}:
\env{align}{
    \mc{G}_\mc{S} =&\ \{g_1, \ldots, g_m\}, \\
    \mc{G}_{\mc{C}(\mc{S})} =&\ \{g_1, \ldots, g_m, h^X_1, \ldots, h^X_{k}, h^Z_1, \ldots, h^Z_{k} \}.
}
Each of $\mc{G}_\mc{S}$ and $\mc{G}_{\mc{C}(\mc{S})}$ is a set of independent operators in $\mc{P}(Q)$; thus
\env{align}{
    |\mc{S}| =&\ 2^{| \mc{G}_\mc{S} |} = 2^m, \\
    |\mc{C}(\mc{S})| =&\ 2^{|\mc{G}_{\mc{C}(\mc{S})}|} = 2^{m+2k} = 2^{|Q| + k}.
}
Each of $\{g_{1\ldots m}\}$, $\{h^X_{1\ldots k}\}$ ,$\{h^Z_{1\ldots k}\}$ is a set of mutually commuting operators in $\mc{P}(Q)$.
In addition, the $g$'s commute with the $h^X$'s as well as with the $h^Z$'s; and $h^X_i h^Z_j = (-1)^{\delta_{ij}} h^Z_j h^X_i$.
The $h$ operators can be thought of the so-called ``representative logical $X$- and $Z$-operators''.

Next, we construct a purification of the state $\rho_Q(\mc{S})$.
Let $R$ be a system of $k$ qubits, and let $\widetilde{\mc{S}} \subseteq \mc{P}(QR)$ be generated by the following set $\widetilde{\mc{G}}$, obtained from $\mc{G}_{\mc{C}(\mc{S})}$ by ``extending'' its elements to $QR$,
\env{align}{
    \label{eq:tilde_G_def}
    &\, \widetilde{\mc{G}} = \ld \(g_j\)_Q \otimes \mathbbm{1}_{R} \Big| j =  1\ldots m \rd %\nn
    %&\hspace{.4in} 
    \cup 
    \ld \(h_j^X\)_Q \otimes \(X_j\)_R \Big| j =  1\ldots k \rd %\nn
    %&\hspace{.4in}
    \cup 
    \ld \(h_j^Z\)_Q \otimes \(Z_j\)_R \Big| j =  1\ldots k \rd,
}
where $(X_j)_R$ is the Pauli $X$-operator on the $j$-th qubit of $R$; and similarly for $(Z_j)_R$.
It is clear that $\widetilde{\mc{G}}$ is a set of independent, mutually commuting elements of $\mc{P}(QR)$, and thus defines a physical state on $QR$,
\env{align}{
    \rho_{QR}(\widetilde{\mc{S}}) = 2^{-|QR|} \sum_{g \in \widetilde{\mc{S}}} g.
}
Since $|\widetilde{\mc{G}}| = |\mc{G}_{\mc{C}(\mc{S})}| = |Q| + k = |QR|$, we have $|\widetilde{\mc{S}}| = 2^{|QR|}$, and from Eq.~\eqref{eq:s_rho_Q_derivation}
\env{align}{
    (\ln 2)^{-1} S(\rho_{QR}(\widetilde{\mc{S}}) ) = 0.
}
Moreover, by construction,
\env{align}{
    \rho_Q(\widetilde{\mc{S}})
    \coloneqq&\, {\rm Tr}_R  \lz  \rho_{QR}(\widetilde{\mc{S}}) \rz \nn
    =&\, 2^{-|QR|} \sum_{g \in \widetilde{\mc{S}}} {\rm Tr}_R (g) \nn
    =&\, 2^{-|Q|} \sum_{g \in \mc{S}} g \nn
    =&\, \rho_Q(\mc{S}).
}
Therefore, $\rho_{QR}(\widetilde{\mc{S}})$ is a purification of $\rho_Q(\mc{S})$ on $QR$, as claimed.

On the other hand, let us compute the reduced density matrix on $R$,
\env{align}{
& \rho_R(\widetilde{\mc{S}}) \nn
    =&\, {\rm Tr}_Q  \lz  \rho_{QR}(\widetilde{\mc{S}}) \rz \nn
    =&\, 2^{-|QR|} \sum_{g \in \widetilde{\mc{S}}} {\rm Tr}_Q (g) \nn
    =&\, 2^{-|R|} \mathbbm{1}_R,
}
i.e. the maximally-mixed state on $R$, as expected.
Thus we have
\env{align}{
    (\ln 2)^{-1} S(\rho_{R}(\widetilde{\mc{S}})) = |R| = k.
}

It is easy to verify that for $A \subseteq Q$ and $\ovl{A} \coloneqq Q - A$,
\env{align}{
    \mathrm{proj}_{\ovl{A} }(\mc{C}(\mc{S})) = \mathrm{proj}_{\ovl{A} }(\widetilde{\mc{S}}),
}
where $\mathrm{proj}_{\ovl{A}}$ on the LHS is understood as from $\mc{P}(Q)$ to $\mc{P}(\ovl{A})$, and that on the RHS from $\mc{P}(QR)$ to $\mc{P}(\ovl{A})$.
Thus, using Eq.~\eqref{eq:s_rho_A_derivation}, but now for $AR \subseteq QR$, $\ovl{A} = Q - A = QR - AR$, and $\widetilde{\mc{S}}$, we have
\env{align}{
    %& \log_2 | \mathrm{proj}_{\ovl{A} }(\mc{C}(\mc{S})) | \nn
    %=& \log_2 | \mathrm{proj}_{\ovl{A} }(\widetilde{\mc{S}}) | \nn
    %=& S\(\rho_{QR - \ovl{A}}(\widetilde{\mc{S}}) \) - (|QR| - |\ovl{A}|) + \log_2 |\widetilde{\mc{S}}| \nn
    %=& (\ln 2)^{-1}S(\rho_{AR}(\widetilde{\mc{S}}) ) - |AR| + \log_2 |\widetilde{\mc{S}}|.
    & (\ln 2)^{-1}S(\rho_{AR}(\widetilde{\mc{S}}) ) \nn
    =& |AR| - \log_2 |\widetilde{\mc{S}}| + \log_2 | \mathrm{proj}_{\ovl{A} }(\widetilde{\mc{S}}) | \nn
    =& |AR| - \log_2 |\widetilde{\mc{S}}| + \log_2 | \mathrm{proj}_{\ovl{A} }(\mc{C}(\mc{S})) |.
}
Combining this equation with Eqs.~(\ref{eq:s_rho_A_derivation}, \ref{eq:size_of_LA}), we have (compare Eq.~\eqref{eq:ell_A_I_AR})
%\env{widetext}{
\env{align}{
    &\ell_A \nn
    =& \log_2 |\mc{L}_A| \nn
    =& \log_2 |\mc{C}(\mc{S})| - \log_2 |\mc{S}| +  \log_2 | \mathrm{proj}_{\ovl{A} }(\mc{S}) | - \log_2 | \mathrm{proj}_{\ovl{A} }(\mc{C}(\mc{S})) | \nn
    =& \log_2 |\mc{C}(\mc{S})| - \log_2 |\mc{S}| %\nn
    %&\ 
    + \Big[ (\ln 2)^{-1} S(\rho_A(\mc{S})) - |A| + \log_2 |\mc{S}| \Big] %\nn
    %&\
    - \lz (\ln 2)^{-1} S(\rho_{AR}(\widetilde{\mc{S}})) - |AR| + \log_2 |\widetilde{\mc{S}}| \rz \nn
    =& \Big[ \log_2 |\mc{C}(\mc{S})| - \log_2 |\widetilde{\mc{S}}| \Big]
    + \Big[ \log_2 |\mc{S}| - \log_2 |\mc{S}| \Big] %\nn
    %&\
    +
    \Big[
    (\ln 2)^{-1} S(\rho_A(\mc{S})) - |A| - (\ln 2)^{-1} S(\rho_{AR}(\widetilde{\mc{S}})) + |AR|
    \Big]
    \nn
    =&\, (\ln 2)^{-1} S(\rho_A(\mc{S})) - (\ln 2)^{-1} S(\rho_{AR}(\widetilde{\mc{S}})) + |R| \nn
    =&\, (\ln 2)^{-1} \lz S(\rho_A(\widetilde{\mc{S}})) - S(\rho_{AR}(\widetilde{\mc{S}})) + 
    S(\rho_{R}(\widetilde{\mc{S}})) \rz \nn
    =&\, (\ln 2)^{-1} I_{A, R}.
}
%}
Thus, we have proven the result stated in Sec.~\ref{sec:theorem}, by constructing a particular purification of $\rho_Q(\mc{S})$ using a particular generating set of $\mc{C}(\mc{S})$.
But this choice is really arbitrary, and there is no surprise that it should work.
In fact, any purification of $\rho_Q(\mc{S})$ on $QR$ with $|R| = k$ has a generating set of the form in Eq.~\eqref{eq:tilde_G_def}, and thus gives a generating set of $\mc{C}(\mc{S})$.

\section{Capillary-wave theory calculations \label{app:cw}}

We compute within capillary-wave theory the free energies of two types of domain walls: those with pinned endpoints, as in Fig.~\ref{fig:bc_ab}(b); and those with free endpoints that wrap around the ``waist'' of the circuit, as in Fig.~\ref{fig:bc_waist}.

\subsection{Domain walls with pinned endpoints}

For the case in Fig.~\ref{fig:bc_ab}(b), we have (compare Eq.~\eqref{eq:Z_y_x_Gaussian})
\env{align}{
    & F_{\rm CW}(A) \\
    =& -\ln \int \mc{D}[y(x)] \exp \lz -\beta \sigma \int_{x_1}^{x_2} dx \sqrt{1 + \( \pd_x y \)^2} \rz,
}
where the functional integral over $y(x)$ is over the following class of ``height functions'',
\env{align}{
    y : [x_1, x_2] \to&\ [-T, 0], \nn
        x \mapsto&\ y(x),
    %y(x), \text{where } x \in [x_1, x_2] \text{ and } y(x_1) = y(x_2) = 0.
}
with the additional constraint that the endpoints are ``pinned'', $y(x_1) = y(x_2) = 0$.
To regularize the path integral, we will however take $y(x_1) = y(x_2) = \epsilon$ to be a small constant, which can be understood as the lattice spacing. 

We first expand the square root,
\env{align}{
    \sqrt{1 + \( \pd_x y \)^2} = 1 + \frac{1}{2} \( \pd_x y \)^2 + O\(\( \pd_x y \)^4\),
}
and neglect quartic and higher order terms in $\( \pd_x y \)$; these are irrelevant under a renormalization group transformation.
Thus we have a Gaussian theory,
\env{align}{
    \label{eq:B5}
    &  F_{\rm CW}(A) \nn
    =& -\ln \int \mc{D}[y(x)] \exp \lz -\beta \sigma \int_{x_1}^{x_2} dx \(1 + \frac{1}{2} \( \pd_x y \)^2 \) \rz \nn
    =&\, \beta \sigma |A| - \ln \int \mc{D}[y(x)] \exp \lz -\frac{\beta \sigma}{2} \int_{x_1}^{x_2} dx \( \pd_x y \)^2 \rz.
}
The second term in this equation is the summation over all admissible configurations of paths/height functions $y(x)$, and can be viewed as a ``random walk'' with ``diffusion constant'' $(\beta \sigma)^{-1}$.
It is thus regarded as the ``thermal entropy'' of transverse fluctuations of the domain walls.
\YL{The magnitude of the fluctuation scales with $x_{12}$ identically to that of a random walker, and can, for example, be quantified by the following quantity,
\env{align}{
    \label{eq:B6}
    \sqrt{\avg{\lz y\(\ovl{x}\) - y(x_1) \rz^2}} \propto \sqrt{\frac{|A|}{\beta \sigma}},
}
where $\ovl{x} \coloneqq (x_1 + x_2)/2$.}

In the following we will, for convenience, treat the path integral in Eq.~\eqref{eq:B5} as a quantum mechanical transition amplitude, \YL{from which Eq.~\eqref{eq:B6} can also be deduced}. However, we note that there are other ways to evaluate this integral, e.g. by solving the diffusion equation \YL{subject to the constraint $y(x) \in [-T, 0]$}.

We now ``quantize'' the path integral, with the spatial direction $x$ viewed as ``imaginary time''. %, and the transverse $y$ direction viewed as ``space''.
We then have an imaginary time path integral of a free quantum particle with mass $\beta \sigma$, confined within a potential well $y \in [-T, 0]$,
\env{align}{
\label{eq:exp_H_tau_pinned}
%&P(y_f, x_f; y_i, x_i) \nn
%=
& \exp\lz -F_{\rm CW}(A) + \beta \sigma |A| \rz \nn
=& \int_{y(x) \in [-T, 0], y(x_1) = y(x_2) = \epsilon} \mc{D}[y(x)] \exp \lz -\frac{\beta \sigma}{2} \int_{x_1}^{x_2} dx \( \pd_x y \)^2 \rz \nn
%=& \int_{y(x) \in [-T, 0], y(x_1) = y(x_2) = \epsilon} \mc{D}[y(x)] \exp \lz -\frac{M}{2} \int_{x_1}^{x_2} dx \( \pd_x y \)^2 \rz \nn
=&
\left\langle y(x_2) \Big\vert \exp \lz -\hat{H} x_{12} \rz
\Big\vert
y(x_1)
\right\rangle,
}
where the Hamiltonian is that of a ``particle in box'' problem,
\env{align}{
    \hat{H} = \frac{\hat{p}_y^2}{2 M} + V(\hat{y}),
    \text{ where } V(y) = \env{cases}{ 0, &-T \le y \le 0; \\
    \infty, &\text{otherwise}.
    }
    %\quad \lz \hat{p}_y, \hat{y} \rz = -i,
}
The eigenstates and their corresponding energies are
\env{align}{
    \phi_n(y) =& \braket{y}{n} = \sqrt{\frac{2}{T}} \sin\( \frac{n \pi y}{T}\), \quad y \in [-T, 0], \nn
    E_n =&\ \frac{1}{2\beta \sigma} \(\frac{n\pi}{T}\)^2, \quad n = 1, 2, 3, \ldots
}
We expand Eq.~\eqref{eq:exp_H_tau_pinned} in the eigenbasis,
\env{align}{
    & \exp\lz -F_{\rm CW}(A) + \beta \sigma |A| \rz \nn
    =& \sum_{n=1}^\infty \braket{y(x_2)}{n} \braket{n}{y(x_1)} \exp \lz -E_n x_{12} \rz \nn
    =&\,\frac{2}{T} \sum_{n=1}^\infty \sin^2 \(\frac{n \pi \epsilon}{T} \) \exp \lz - \frac{1}{2\beta \sigma} \(\frac{n \pi \sqrt{x_{12}}}{T}\)^2 \rz.
}
When $\pi \sqrt{x_{12}} / T \ll 1$, we may approximate the summation with the following integral over $u = \frac{n \pi \sqrt{x_{12}}}{T}$,
\env{align}{
    & \exp\lz -F_{\rm CW}(A) + \beta \sigma |A| \rz \nn
    \approx&\frac{2}{T} \int_{\pi \sqrt{x_{12}} / T}^\infty \frac{T du}{\pi \sqrt{x_{12}}} \sin^2\(\frac{u \epsilon}{\sqrt{x_{12}}}\) \exp(-\frac{u^2}{2\beta \sigma}) \nn
    \approx&\frac{2}{\pi \sqrt{x_{12}}} \int_0^\infty du \sin^2\(\frac{u \epsilon}{\sqrt{x_{12}}}\) \exp(-\frac{u^2}{2\beta \sigma}) \nn
    \approx&\frac{2 \epsilon^2}{\pi (x_{12})^{3/2}} \int_0^\infty du\, u^2 \exp(-\frac{u^2}{2\beta \sigma}) \nn
    =&\frac{2 \epsilon^2}{\pi |A|^{3/2}} \int_0^\infty du\, u^2 \exp(-\frac{u^2}{2\beta \sigma}) \nn
    =&\, \sqrt{\frac{2}{\pi}} \epsilon^2 (\beta \sigma)^{3/2} |A|^{-3/2},
}
and thus (compare Eq.~\eqref{eq:s_rho_A_small})
\env{align}{
    \label{eq:s_rho_A_small_appB}
    &\, F_{\rm CW}(A) \nn
    =&\, \beta \sigma |A| + \frac{3}{2} \ln |A| + \mathrm{const.}, \text{ when } \sqrt{|A|} \ll T.
}
In arriving at this result, we made the following replacement in the integrand
\env{align}{
    \sin^2\(\frac{u \epsilon}{\sqrt{x_{12}}}\) \exp(-\frac{u^2}{2\beta \sigma}) 
    \quad \to \quad 
    \(\frac{u \epsilon}{\sqrt{x_{12}}}\)^2 \exp(-\frac{u^2}{2\beta \sigma});
}
this is valid when
\env{align}{
    \frac{\epsilon \sqrt{\beta \sigma} }{\sqrt{x_{12}}} \ll 1 
    \quad \Leftrightarrow \quad \epsilon \ll \sqrt{\frac{|A|}{\beta \sigma}}.
}
Physically, it means that the temperature cannot be too low, so that the transverse fluctuation of the domain wall is large compared to the ``lattice spacing'', $\epsilon$.
This is consistent with the $p=0$ limit of the circuit (now without measurements), corresponding to the zero-temperature limit of capillary-wave theory, where the subleading logarithmic term is absent in the entanglement entropy (see discussions near footnote~\ref{fn:adam_tianci}).

\subsection{``Waist'' domain walls}

As shown in Fig.~\ref{fig:bc_waist}, the ``waist'' domain wall for open b.c. has two independent free endpoints; whereas for periodic b.c. the two endpoints must coincide, but otherwise free.

In the case of open b.c., let $y_1 = y(x_1 = 0)$, and $y_2 = y(x_2 = L)$.
The analog to Eq.~\eqref{eq:exp_H_tau_pinned} reads
%\env{widetext}{
\env{align}{
\label{eq:exp_H_tau_obc}
%&P(y_f, x_f; y_i, x_i) \nn
%=
& \exp\lz -F_{\rm CW}(Q) + \beta \sigma |Q| \rz \nn
=& \int_{y(x) \in [-T, 0]} \mc{D}[y(x)] \exp \lz -\frac{\beta \sigma}{2} \int_{0}^{L} dx \( \pd_x y \)^2 \rz \nn
%=& \int_{y(x) \in [-T, 0], y(x_1) = y(x_2) = 0} \mc{D}[y(x)] \exp \lz -\frac{M}{2} \int_{x_1}^{x_2} dx \( \pd_x y \)^2 \rz \nn
=&
\int_{-T}^0 dy_1 \int_{-T}^0 dy_2 \left\langle y_2 \Big\vert \exp \lz -\hat{H} L \rz
\Big\vert
y_1
\right\rangle \nn
=& \int_{-T}^0 dy_1 \int_{-T}^0 dy_2 \sum_{n=1}^\infty \braket{y_2}{n} \braket{n}{y_1} \exp \lz -E_n L \rz \nn
=&\,\frac{2}{T} \int_{-T}^0 dy_1 \int_{-T}^0 dy_2 %\nn
%&\
\sum_{n=1}^\infty \sin \(\frac{n \pi y_1}{T} \)  \sin \(\frac{n \pi y_2}{T} \)  \exp \lz - \frac{1}{2\beta\sigma} \(\frac{n \pi \sqrt{L}}{T}\)^2 \rz \nn
=&\sum_{n \text{ odd}} \frac{8T}{n^2 \pi^2} \exp \lz - \frac{1}{2\beta\sigma} \(\frac{n \pi \sqrt{L}}{T}\)^2 \rz \nn
\approx&\, \frac{1}{2} \int_{\pi \sqrt{L}/T}^{\infty} \frac{T du}{\pi \sqrt{L}} \frac{8L}{T u^2} \exp \lz -\frac{u^2}{2\beta\sigma} \rz  \nn
=&\, \frac{4\sqrt{L}}{\pi} \int_{\pi \sqrt{L}/T}^{\infty} \, du\, u^{-2} \exp \lz -\frac{u^2}{2\beta\sigma} \rz  \nn
=&\, \frac{4\sqrt{L}}{\pi} \Bigg\{ \lz -u^{-1} \exp \lz -\frac{u^2}{2\beta\sigma} \rz \rz \Bigg|_{\pi \sqrt{L}/T}^{\infty} %\nn
%&\hspace{.6in}
-  \int_{\pi \sqrt{L}/T}^{\infty} \, du\, (-u^{-1}) \(-\frac{u}{\beta\sigma}\) \exp \lz -\frac{u^2}{2\beta\sigma} \rz  \Bigg\} \nn
\approx&\, \frac{4}{\pi^2} T,
}
%}
where we assumed $T \gg \sqrt{L}$ throughout.

For periodic b.c., letting $y = y(x_1 = 0) = y(x_2 = L)$, we have
%\env{widetext}{
\env{align}{
\label{eq:exp_H_tau_pbc}
%&P(y_f, x_f; y_i, x_i) \nn
%=
& \exp\lz -F_{\rm CW}(Q) + \beta \sigma |Q| \rz \nn
=& \int_{y(x) \in [-T, 0], y(0) = y(L)} \mc{D}[y(x)] \exp \lz -\frac{\beta \sigma}{2} \int_{0}^{L} dx \( \pd_x y \)^2 \rz \nn
%=& \int_{y(x) \in [-T, 0], y(x_1) = y(x_2) = 0} \mc{D}[y(x)] \exp \lz -\frac{M}{2} \int_{x_1}^{x_2} dx \( \pd_x y \)^2 \rz \nn
=&
\int_{-T}^0 dy \left\langle y \Big\vert \exp \lz -\hat{H} L \rz
\Big\vert
y
\right\rangle \nn
=& \int_{-T}^0 dy \sum_{n=1}^\infty \braket{y}{n} \braket{n}{y} \exp \lz -E_n L \rz \nn
%=&\,\frac{2}{T} \int_{-T}^0 dy \sum_{n=1}^\infty \sin^2 \(\frac{n \pi y}{T} \) \exp \lz - \frac{1}{2M} \(\frac{n \pi \sqrt{L}}{T}\)^2 \rz \nn
=&\, \sum_{n=1}^\infty \exp \lz - \frac{1}{2\beta\sigma} \(\frac{n \pi \sqrt{L}}{T}\)^2 \rz \nn
%=&\sum_{n \text{ odd}} \frac{8T}{n^2 \pi^2} \exp \lz - \frac{1}{2M} \(\frac{n \pi \sqrt{L}}{T}\)^2 \rz \nn
\approx&\, \int_{\pi \sqrt{L}/T}^{\infty} \frac{T du}{\pi \sqrt{L}} %\frac{8L}{T u^2}
\exp \lz -\frac{u^2}{2\beta\sigma} \rz  \nn
\approx&\, \frac{T}{\pi \sqrt{L}} \int_{0}^{\infty} du\, %\frac{8L}{T u^2}
\exp \lz -\frac{u^2}{2\beta\sigma} \rz  \nn
%=&\, \frac{4\sqrt{L}}{\pi} \int_{\pi \sqrt{L}/T}^{\infty} \, du\, u^{-2} \exp \lz -\frac{u^2}{2M} \rz  \nn
%=&\, \frac{4\sqrt{L}}{\pi} \ld \lz -u^{-1} \exp \lz -\frac{u^2}{2M} \rz \rz \Bigg|_{\pi \sqrt{L}/T}^{\infty} -  \int_{\pi \sqrt{L}/T}^{\infty} \, du\, (-u^{-1}) \(-\frac{u}{M}\) \exp \lz -\frac{u^2}{2M} \rz  \rd \nn
%\approx&\, \frac{4}{\pi^2} T,
=& \sqrt{\frac{\beta \sigma}{2\pi}} \frac{T}{\sqrt L},
}
%}
where we again assumed $T \gg \sqrt{L}$ throughout.

Summarizing, we have (compare Eq.~\eqref{eq:s_rho_Q})
\env{align}{
&\, F_{\rm CW}(Q) \nn
=& \env{cases}{
        \beta \sigma L - \ln T + \mathrm{const.}, & \text{open b.c.}\\
        \beta \sigma L - \ln \frac{T}{\sqrt{L}} + \mathrm{const.},& \text{periodic b.c.}
    }
    \text{ when } T \gg \sqrt{L}.
}
Similarly to domain walls with pinned endpoints, the subleading logarithmic term can again be understood as coming from thermal entropies of transverse fluctuations.
The $\ln \sqrt{L}$ difference is the extra endpoint entropy in open b.c..

\end{widetext}

%---------------------------------------------------------------------------------------------------------
\bibliography{refs}

\end{document}